\documentclass[11pt]{article}

\usepackage[preprint]{acl}

\usepackage{times}
\usepackage{latexsym}

\usepackage[T1]{fontenc}
\usepackage[utf8]{inputenc}

\usepackage{microtype}

\usepackage{inconsolata}

\usepackage{graphicx}

\usepackage{booktabs} 
\usepackage{multirow}
\usepackage{amsmath}
\usepackage{amsfonts}
\usepackage{tabularx}   % 自动列宽
\usepackage{float} 
\usepackage{listings} % 用于读取代码或纯文本文件
\usepackage{subcaption} %

\usepackage[most]{tcolorbox}
\newtcolorbox{promptbox}[1][]{
  colback=gray!5,
  colframe=gray!60,
  coltitle=black,
  fonttitle=\bfseries,
  title={System Prompt},
  sharp corners=south,
  enhanced,
  attach boxed title to top left={yshift=-2mm, xshift=2mm},
  boxed title style={colback=gray!20, boxrule=0pt},
  #1
}
\newtcolorbox{user_promptbox}[1][]{
  colback=gray!5,
  colframe=gray!60,
  coltitle=black,
  fonttitle=\bfseries,
  title={User Prompt},
  sharp corners=south,
  enhanced,
  attach boxed title to top left={yshift=-2mm, xshift=2mm},
  boxed title style={colback=gray!20, boxrule=0pt},
  #1
}

\usepackage[table]{xcolor}

\title{SDiaReward: Modeling and Benchmarking Spoken Dialogue \\Rewards with Modality and Colloquialness}
\author{
  \textbf{Jingyu Lu\thanks{\ \ Equal contribution.}},
  \textbf{Yuhan Wang\footnotemark[1]},
  \textbf{Fan Zhuo\footnotemark[1]},
  \textbf{Xize Cheng},
  \textbf{Changhao Pan}, \\
  \textbf{Xueyi Pu},
  \textbf{Yifu Chen},
  \textbf{Chenyuhao Wen},
  \textbf{Tianle Liang},
  \textbf{Zhou Zhao\thanks{\ \ Corresponding author.}} \\
  Zhejiang University \\
  \texttt{\{lujingyu, zhaozhou\}@zju.edu.cn}
}

\begin{document}
\maketitle
\begin{abstract}
The rapid evolution of end-to-end spoken dialogue systems demands transcending mere textual semantics to incorporate paralinguistic nuances and the spontaneous nature of human conversation. However, current methods struggle with two critical gaps: the \textit{modality gap}, involving prosody and emotion, and the \textit{colloquialness gap}, distinguishing written scripts from natural speech. To address these challenges, we introduce \textbf{SDiaReward}, an end-to-end multi-turn reward model trained on \textbf{SDiaReward-Dataset}, a novel collection of episode-level preference pairs explicitly targeting these gaps. It operates directly on full multi-turn speech episodes and is optimized with pairwise preference supervision, enabling joint assessment of modality and colloquialness in a single evaluator. We further establish \textbf{ESDR-Bench}, a stratified benchmark for robust episode-level evaluation. Experiments demonstrate that SDiaReward achieves state-of-the-art pairwise preference accuracy, significantly outperforming general-purpose audio LLMs. Further analysis suggests that SDiaReward captures relative conversational expressiveness beyond superficial synthesis cues, improving generalization across domains and recording conditions.
Code, data, and demos are available at \url{https://github.com/MM-Speech/SDiaReward/}.
% The demo page can be found at \url{https://sdiareward.github.io/}.
\end{abstract}

\section{Introduction}

\begin{figure}[t]
  \includegraphics[width=\columnwidth]{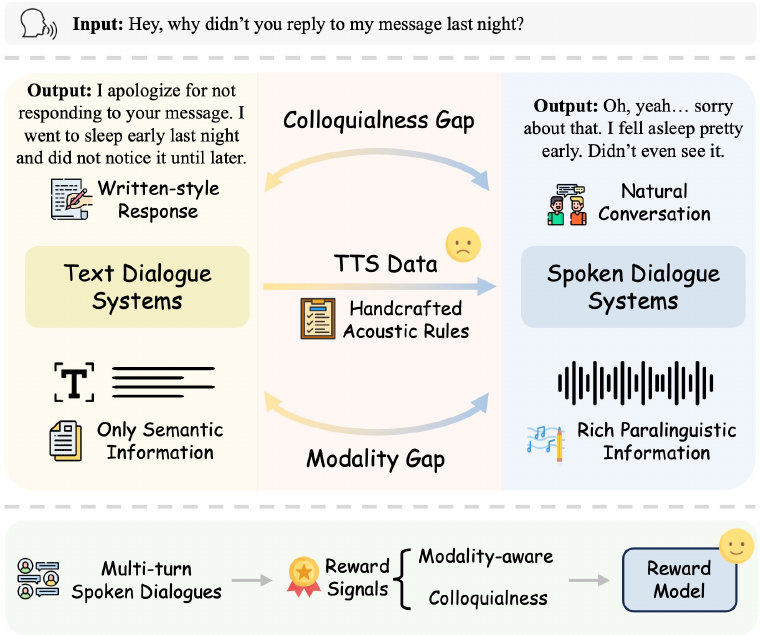}
  \vspace{-6pt}
  \caption{Challenges in spoken dialogue and our proposed framework. Text-based systems face \textbf{modality} (prosody/emotion) and \textbf{colloquialness} (style) gaps. Unlike rule-based methods, our end-to-end Reward Model learns these features from multi-turn dialogues via data-driven preference signals.}
  \label{fig:head}
  \vspace{-8pt}
\end{figure}

Large Language models (LLMs) have driven rapid progress in text-based dialogue systems~\cite{zhao2023survey}, and recent efforts have begun to extend these capabilities to end-to-end spoken dialogue systems that directly perceive and generate speech~\cite{zhang-etal-2023-speechgpt, xie2024mini,defossez2024moshi}. Spoken dialogue promises a more natural interface for human--AI interaction, yet it also raises a fundamental question: how should we reliably evaluate and optimize spoken dialogue behaviors? In practice, progress in text~\cite{ouyang2022training, cai2024internlm2, jiang-etal-2023-llm,zheng2023judging} and vision~\cite{wang2025unified,zang2025internlm} has been strongly enabled by reward modeling~\cite{zhong2025comprehensive} and preference learning~\cite{christiano2017deep, rafailov2023direct}, which provide scalable supervision for alignment, reranking, and reinforcement learning. However, reliable reward modeling and evaluation for end-to-end spoken dialogue remains underexplored.

A key reason is that moving from text dialogue to spoken dialogue exposes two gaps that complicate reward design and evaluation.
\emph{i) Modality gap:} speech carries paralinguistic information such as prosody, emotion, and channel conditions. These elements strongly influence human preference yet remain invisible to text-based evaluators.
\emph{ii) Colloquialness gap:} written-style responses produced by text-optimized systems are often well-formed but sound overly scripted when spoken, while natural conversation prefers brevity, fragmentation, discourse markers, and interactional cues~\cite{yan-etal-2025-uro, chen2024voicebench}.
Crucially, standard general-purpose evaluators often exhibit ``modality blindness''---failing to distinguish between natural human speech and synthesized artifacts when semantic content is identical. Instead of relying on rigid acoustic rules, we argue for a \textbf{data-driven paradigm} where reward signals for paralinguistic fidelity and interactional spontaneity are implicitly learned from large-scale preference comparisons.

In this work, we conduct a benchmark-driven study of spoken dialogue reward modeling and evaluation.
We formulate pairwise preference supervision for multi-turn spoken dialogues, and decompose reward signals into two aspects: (i) a modality-aware component that evaluates content adequacy, dialogue coherence, and spoken naturalness, and (ii) a colloquialness component that captures stylistic and interactional properties of spontaneous speech.
We then establish \textsc{ESDR-Bench}, a carefully stratified benchmark designed with multi-dimensional annotations to ensure distributional diversity. This enables rigorous assessment of model generalization beyond standard random splits.
Experiments demonstrate that our data-driven model achieves state-of-the-art pairwise preference accuracy, significantly outperforming general-purpose Audio LLMs which struggle with the modality gap.
Further analysis suggests that instead of merely detecting artifacts, our model captures relative conversational expressiveness, implicitly calibrating preference rankings within diverse acoustic domains.
\\Our contributions are threefold:
\begin{itemize}
    \item \textbf{Dataset.} We construct a Spoken Dialogue Reward Dataset (\textsc{SDiaReward-Dataset}) containing 11k preference pairs (200 hours of paired speech) for training spoken dialogue reward models. The full dataset will be released openly following necessary de-identification and ethics clearance.
    % with a public release planned in the near future following additional ethics review and data de-identification procedures.
    \item \textbf{Reward Modeling Framework.} We introduce an end-to-end spoken dialogue reward modeling framework in a pairwise setting and decompose evaluation into modality-aware and colloquialness rewards for multi-turn spoken dialogues.
    \item \textbf{Benchmark \& Analysis.} We construct an episode-level spoken dialogue reward benchmark (\textsc{ESDR-Bench}) with multi-dimensional annotations. Based on this, we provide an empirical analysis demonstrating the superiority of specialized data-driven reward modeling over generalist judges, offering practical insights for reliable spoken dialogue alignment.
\end{itemize}

\section{Related Work}
\begin{figure*}[t]
\centering
\includegraphics[width=\linewidth]{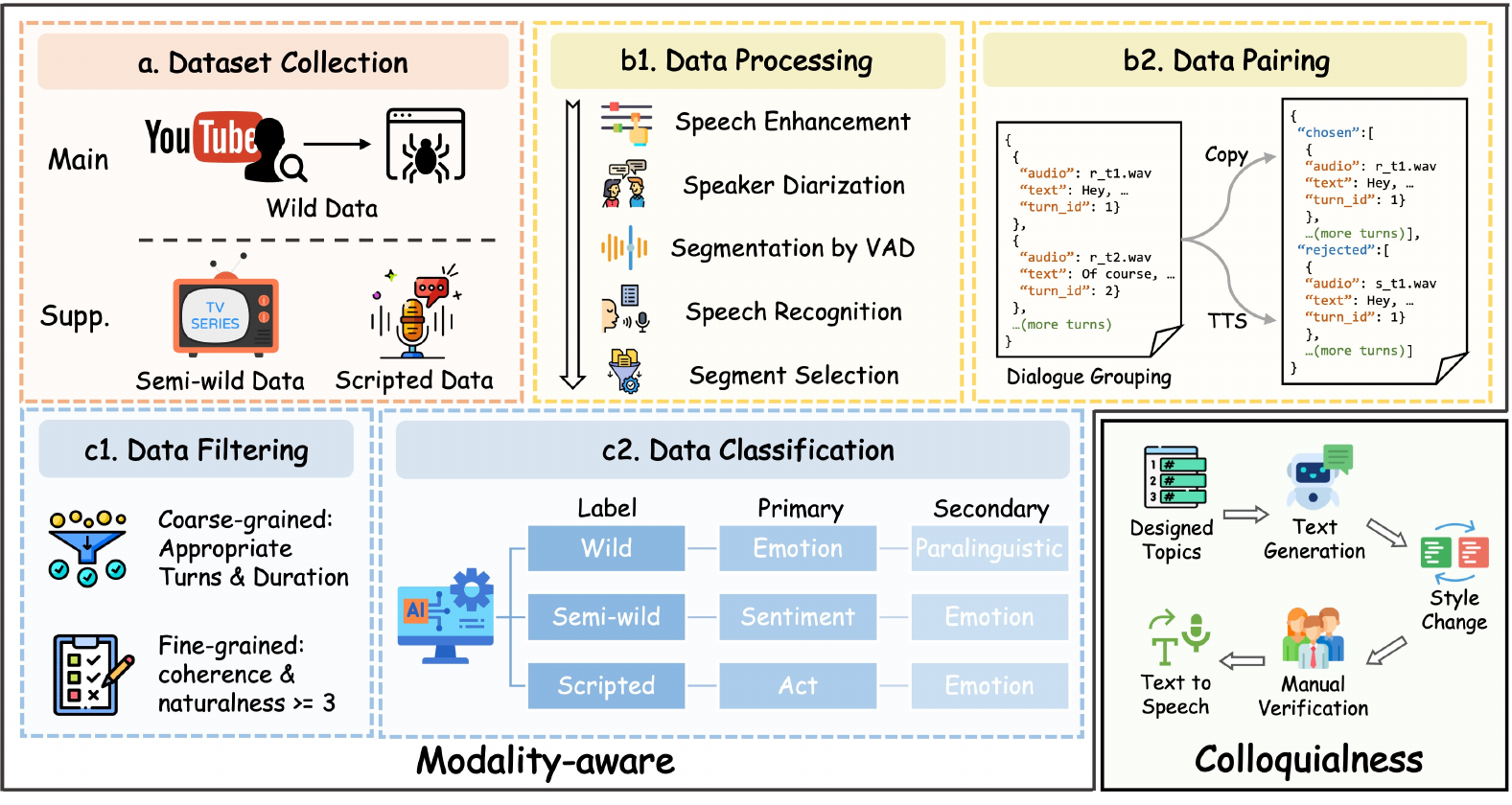}
\vspace{-6pt}
\caption{Overview of dataset construction. (a) \textbf{Collection:} We collect wild conversational audio (main) along with semi-wild/scripted data. (b1--b2) \textbf{Processing \& Pairing:} We process audio into speaker-aware turns and group them into dialogues. We then construct two types of pairs: \textit{modality-aware pairs} (center) via real vs.\ TTS audio , and \textit{colloquialness pairs} (bottom right) via text-style vs.\ spoken-style generation and style change. (c1--c2) \textbf{Post-processing:} We filter episodes and attach hierarchical metadata (emotion, sentiment, act) for benchmark stratification. The detailed data processing pipeline can be found in Appendix ~\ref{sec:appendix_dataset}.}
\label{fig:data_pipeline}
\vspace{-8pt}
\end{figure*}

\paragraph{End-to-End Spoken Dialogue and Alignment}
The recent transition from cascaded pipelines to end-to-end spoken dialogue systems marks a significant shift in conversational AI~\cite{ji2024wavchat}, driven by advancements in acoustic tokenization~\cite{ji2024wavtokenizer}, scalable data synthesis~\cite{cheng2025omnichat}, and audio-integrated retrieval~\cite{chen2025wavrag}, enabling models to integrate acoustic perception and speech generation within a unified framework~\cite{zhang-etal-2023-speechgpt,xie2024mini,defossez2024moshi,xu2025qwen2}. While these systems promise enhanced interactivity and paralinguistic expressiveness, they present unique challenges for evaluation and optimization. Unlike text-based dialogue, spoken outputs must satisfy not only semantic adequacy but also prosodic naturalness and interactional spontaneity. In the textual domain, reward modeling has established itself as a cornerstone for alignment, employing techniques such as reinforcement learning from human feedback and direct preference optimization to steer model behaviors~\cite{christiano2017deep,ouyang2022training,rafailov2023direct}. However, extending these paradigms to the auditory domain remains non-trivial. While recent adaptive post-training frameworks have shown promise in aligning these models for enhanced expressiveness~\cite{chen2026wavalignenhancingintelligenceexpressiveness}, text-centric reward models inherently overlook the modality gap, while traditional automatic metrics fail to account for the colloquial nuances and long-range coherence required in spontaneous, multi-turn interaction.

\paragraph{Multimodal and Speech Reward Modeling}
As reward modeling expands beyond text, substantial research has focused on multimodal generation and understanding, accompanied by rigorous benchmarking efforts to ensure reliability and mitigate bias~\cite{xu2023imagereward,yu2024rlhf,wang2025unified,lambert2025rewardbench,liu2024rm}. In the speech domain, while recent benchmarks have begun assessing reasoning, colloquialism, and non-verbal understanding in spoken dialogues~\cite{cheng2025voxdialogue, li2026wavbench}, preference modeling itself remains relatively under-explored. Existing approaches such as SpeechJudge~\cite{zhang2025speechjudge} primarily target single-turn text-to-speech quality assessment. While emerging generative reward models have started to address semantic and turn-taking robustness in interactive systems~\cite{chen2026dualaxisgenerativerewardmodel}, other recent initiatives, including ParaS2S~\cite{yang2025paras2s} and WavReward~\cite{ji2025wavreward}, incorporate paralinguistic signals but often depend on manually defined acoustic features or rules, which may lack the flexibility to generalize to the diversity of "wild" conversational data. Distinct from these methods, our framework addresses these limitations by establishing a holistic, episode-level reward model. We aim to learn both acoustic plausibility and conversational colloquialness directly from data, bypassing the reliance on handcrafted engineering and enabling general evaluation of multi-turn spoken dialogues.

\section{Dataset and Benchmark}
\label{sec:data}

We introduce \textbf{SDiaReward-Dataset}, a large-scale corpus specifically constructed to enable episode-level reward modeling for spoken dialogue. The dataset addresses two fundamental gaps that hinder current evaluation methods: the \emph{modality gap}, which stems from the loss of paralinguistic cues such as prosody and emotion in standard synthesis, and the \emph{colloquialness gap}, which arises from the stylistic divergence between rigid written scripts and spontaneous natural speech.
To bridge these gaps, we curate contrastive dialogue pairs that provide supervision signals for both dimensions. The \textit{modality-aware} subset juxtaposes real human speech with synthesized counterparts, training the model to discern authentic paralinguistic fidelity from synthesis artifacts while controlling for linguistic content. Complementarily, the \textit{colloquialness} subset contrasts formal written-style interactions with spoken-style rewrites under consistent acoustic conditions, targeting the optimization of conversational flow and interactional spontaneity.
The resulting corpus comprises approximately 13k pairwise samples (Table~\ref{tab:data-stats}), from which we establish our stratified evaluation benchmark, \textbf{ESDR-Bench}.

\begin{table}[t]
\centering
\small 
\caption{Statistics of the dataset. We categorize data by modality types and colloquialness. The unit is the pairs of dialogue.}
\begin{tabular}{lrrr} 
\toprule
\textbf{Category} & \textbf{Train} & \textbf{Val} & \textbf{Total} \\ 
\midrule
\textit{Modality} & & & \\
\quad Wild       & 6,879  & 824   & 7,703  \\
\quad Semi-Wild  & 309    & 186   & 495   \\
\quad Scripted   & 2,192  & 466   & 2,658   \\
\midrule
\textit{Colloquialness}  & 2,250  & 250   & 2,500  \\ 
\midrule
\textbf{Total All} & \textbf{11,630} & \textbf{1,726} & \textbf{13,356} \\ 
\bottomrule
\end{tabular}
\label{tab:data-stats}
\vspace{-11pt}
\end{table}

% \subsection{Design Goals and Dataset Overview}
% To study spoken dialogue rewards, we curate data that explicitly targets the modality and colloquialness gaps. 
% The modality-aware subset pairs real-world spoken dialogue episodes with synthesized counterparts, encouraging reward signals that are sensitive to prosody, expressiveness, and channel conditions while controlling for dialogue content.
% The colloquialness subset focuses on style: it pairs written-style dialogues with spoken-style dialogues, while keeping the speech synthesis condition consistent to reduce acoustic confounds.
% Both subsets are merged to train a single reward model that outputs a scalar reward for a multi-turn spoken dialogue episode.
% Figure~\ref{fig:data_pipeline} summarizes the overall pipeline from collection to benchmark construction.

\subsection{Construction Pipeline}

\paragraph{Real-world Audio Collection.}
We implement a systematic pipeline designed to transform unconstrained web audio into high-quality, structured dialogue episodes. Targeting the \textit{Wild} condition, we crawl long-form conversational content from curated YouTube domains to maintain thematic consistency. To address the inherent acoustic variability of web sources, the data undergoes a multi-stage processing chain that includes speech enhancement for noise reduction, neural speaker diarization to disentangle overlapping speech, and VAD-guided segmentation aligned with ASR transcripts. Crucially, we preserve the sequential dependencies of the original recordings by grouping segments into continuous multi-turn episodes. This structure enables the model to capture global conversational dynamics and context-dependent prosody rather than focusing solely on isolated utterance quality.

\paragraph{Modality-aware Pairing.}
To construct the modality-aware subset, we juxtapose authentic human speech with synthesized counterparts generated by \texttt{SoulX-Podcast}~\cite{xie2025soulx}. We selected this Dialogue-TTS system for its capacity to maintain multi-turn speaker coherence, ensuring high-fidelity ``hard negatives'' that force the model to discern subtle prosodic naturalness rather than trivial discontinuity artifacts. The human speech sources are stratified into three tiers:
\textbf{1) Wild Data}, spontaneous multi-speaker conversations from YouTube with authentic background noise;
\textbf{2) Semi-wild Data}, derived from MELD~\cite{poria2019meld}, featuring emotionally rich acted dialogues; and
\textbf{3) Scripted Data}, sourced from DailyTalk~\cite{lee2023dailytalk}, representing high-fidelity studio recordings.
By pairing these diverse sources with dialogue-consistent synthesis, we isolate acoustic realization as the primary discriminative factor, prioritizing paralinguistic naturalness over spectral cleanliness.
We deliberately selected this context-aware Dialogue-TTS system to prevent the model from exploiting severe discontinuity artifacts common in early single-turn systems as shortcuts, ensuring high-fidelity ``hard negatives'' that force the model to discern true paralinguistic naturalness over spectral cleanliness.
% \paragraph{Modality-aware Pairing.}
% To ensure model generalization across diverse acoustic environments, we construct the modality-aware subset by juxtaposing authentic human speech with synthesized counterparts generated by \texttt{SoulX-Podcast}~\cite{xie2025soulx}. The human speech sources are stratified into three tiers to cover a spectrum of spontaneity:
% \textbf{1) Wild Data}, consisting of spontaneous multi-speaker conversations from YouTube that contain authentic background noise and natural disfluencies;
% \textbf{2) Semi-wild Data}, derived from the MELD dataset~\cite{poria2019meld}, which features emotionally rich and acted dialogues from TV series; and
% \textbf{3) Scripted Data}, sourced from DailyTalk~\cite{lee2023dailytalk} to represent high-fidelity studio recordings with strictly controlled linguistic content.
% By pairing these diverse human recordings with dialogue-consistent synthesized speech, we isolate acoustic realization as the primary discriminative factor. This encourages the reward model to prioritize paralinguistic naturalness and expressivity over spectral cleanliness.

\paragraph{Colloquialness Pairing.}
This subset targets the stylistic gap between formal text and spontaneous speech by contrasting written-style dialogues against spoken-style rewrites. We initially design 250 scenarios across 10 domains and employ LLMs to generate multi-turn written-style dialogues. To mitigate potential ``LLM-style'' bias, these scripts are subsequently rewritten into spoken-style versions using fine-grained linguistic constraints. They preserve the original meaning but incorporate natural conversational patterns such as fillers, fragmentation, and discourse markers, which manually confirmed to naturally induce more realistic pause and breath patterns when rendered by TTS. To prevent acoustic quality from confounding the preference signal, we synthesize both the written and spoken versions using the identical TTS configuration. Consequently, the preference labels rely exclusively on the stylistic naturalness of the dialogue flow rather than differences in audio fidelity.
% This subset targets the stylistic gap between formal text and spontaneous speech by contrasting written-style dialogues against spoken-style rewrites. We initially design 250 scenarios across 10 domains and employ LLMs to generate multi-turn written-style dialogues. These scripts are subsequently rewritten into spoken-style versions that preserve the original meaning but incorporate natural conversational patterns such as fillers, fragmentation, and discourse markers. To prevent acoustic quality from confounding the preference signal, we synthesize both the written and spoken versions using the identical TTS configuration. Consequently, the preference labels rely exclusively on the stylistic naturalness of the dialogue flow rather than differences in audio fidelity.

\subsection{Quality Control and ESDR-Bench}
\label{sec:3_2}

\paragraph{Filtering and Annotation.}
To guarantee the reliability of the preference signals, we enforce a rigorous two-stage quality control protocol. Structurally, we limit dialogues to a maximum of 16 turns and restrict individual turn durations to 60 seconds to maintain manageable sequence lengths. Qualitatively, we employ an LLM-based judge to assess episode quality across three dimensions: content adequacy, dialogue coherence, and prosodic naturalness. We discard any episodes that fail to achieve a minimum threshold of 3 out of 5 on the adequacy and coherence scales. To facilitate fine-grained performance analysis, we further enrich the dataset with metadata annotations, including emotion tags and sentiment labels derived from the source material or predicted by auxiliary models.

% \begin{figure}[t]
%     \centering
%     \includegraphics[width=0.92\linewidth]{latex/imgs/dataset_overview.pdf}
%     \vspace{-6pt}
%     \caption{Overview of the ESDR-Bench statistics.}
%     \label{fig:bench_stat}
%     \vspace{-8pt}
% \end{figure}

\paragraph{Benchmark Stratification.}
We establish the \textbf{ESDR-Bench} from the held-out validation split to serve as a robust evaluation standard. A key challenge in benchmark construction is the potential dominance of high-frequency data types such as the Wild subset. To address this imbalance, we implement a stratified sampling strategy based on source and metadata categories. For each fine-grained bucket, we select a balanced set of up to 50 episodes, ensuring that the benchmark provides a distributionally diverse assessment of model generalization rather than being skewed by the underlying data distribution.
Although the collected corpus naturally contains more Wild audio due to availability, ESDR-Bench uses source- and metadata-stratified sampling to prevent high-frequency regimes from dominating evaluation and to better reflect generalization.

\section{Reward Modeling}
\label{sec:method}

\paragraph{Problem Setup.}
We consider a multi-turn spoken dialogue as a sequence of turns $\mathcal{D} = \{(a_t, x_t)\}_{t=1}^{T}$, where $a_t$ denotes the speech audio and $x_t$ represents the corresponding transcript. 
Unlike traditional reward models that focus on isolated turns, our goal is to evaluate the contextual consistency and multimodal alignment of a candidate final turn. 
Given a context $\mathcal{C}=\{(a_t,x_t)\}_{t=1}^{T-1}$ and candidate final turns $y$ , the model outputs scalar rewards $r_\theta(\mathcal{C}, y)$ leveraging complete context information of the conversation.
% the model outputs scalar rewards $r_\theta(\mathcal{C}, y)$.
% We consider a multi-turn spoken dialogue as a sequence of turns
% $\mathcal{D} = \{(a_t, x_t)\}_{t=1}^{T}$,
% where $a_t$ is the speech audio, $x_t$ is the transcript. 
% Our goal is to learn a reward model that can evaluate the consistency of the final spoken response with context in real-world. % 想一下怎么把奖励模型的目标说清楚，围绕模态奖励和口语化奖励
% Given a context $\mathcal{C}=\{(a_t,x_t)\}_{t=1}^{T-1}$ and candidate final turns $y$ ,
% the model outputs scalar rewards $r_\theta(\mathcal{C}, y)$.
\begin{figure}[t]
\centering
\includegraphics[width=0.9\linewidth]{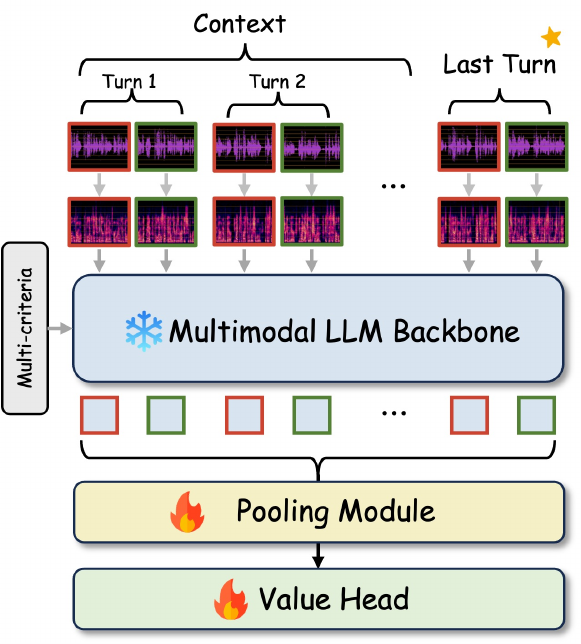}
\vspace{-6pt}
\caption{Architecture of our reward model.}
\label{fig:model}
\vspace{-8pt}
\end{figure}

\paragraph{Model Architecture.}
Existing speech preference models often focus on single-turn TTS evaluation, neglecting the long-range dependency in dialogues. Others rely on handcrafted paralinguistic features, which lack the capacity to capture the nuanced "vibe" of spontaneous speech.
To address these critical limitations in spoken dialogue rewarding, we develop an end-to-end multimodal reward model designed to capture the complex alignment between speech context and response.
We leverage a multimodal LLM backbone to project the interleaved speech-text sequence into a joint embedding space. Let $\mathbf{H} = \{h_1,\ldots,h_L\} \in \mathbb{R}^{L\times d}$ be the hidden representations extracted from the final transformer layer. 
The scalar reward is then computed via a task-specific score head:
\begin{equation}
r_\theta(\mathcal{C},y) = \text{MLP}(\textsc{Pool}(\mathbf{H})),
\end{equation}
where $\textsc{Pool}(\cdot)$ is a pooling operator that aggregates sequence-level information. This architecture bypasses the need for intermediate text-based summarization, allowing the model to directly "hear" the prosodic nuances in the context.
% \paragraph{Model Architecture}
% Existing speech preference models often focus on single-turn TTS evaluation or rely on handcrafted paralinguistic rules, which can miss dialogue grounding and may not scale to the diversity of spontaneous speech.
% In contrast, we model rewards end-to-end on speech dialogues by conditioning on the full multi-turn context.
% Concretely, we use a multimodal LLM backbone to obtain the hidden representations 
% $\mathbf{H} = \{h_1,\ldots,h_L\}$ of the dialogue sequence, where $\mathbf{H} \subseteq \mathbb{R}^{L\times d}$, $d$ is the hidden dimension.
% We then apply a pooling operator $\textsc{Pool}(\cdot)$ to obtain a fixed-dimensional dialogue representation for feeding into the socre head to calculate the reward score:
% \begin{equation}
% r_\theta(\mathcal{C},y) = \text{Head}(z),\quad \text{where}\ z = \textsc{Pool}(\mathbf{H}).
% \end{equation}

\paragraph{Pooling and Robustness}
A practical consideration is how to summarize the sequence representation $\mathbf{H}$ for reward prediction.
We evaluate three standard pooling operators: last-token pooling, mean pooling, and attention pooling.
Empirically, mean pooling provides the most stable optimization behavior across hyperparameters and data mixtures, while attention pooling can achieve high accuracy but exhibits higher sensitivity and may allocate criterion-dependent attention patterns across distributions.
Last-token pooling underperforms in our setting, suggesting that reward-relevant information is distributed across the context and final-turn representations rather than concentrated in a single position.
We defer detailed ablations and quantitative comparisons to \S\ref{sec:exp}.

% \mathcal{L}_{\text{pref}}(\theta)
% = - \mathbb{E}_{(\mathcal{C},y^+,y^-)} \left[
% \log \sigma\left(r_\theta(\mathcal{C},y^+) - r_\theta(\mathcal{C},y^-)\right)
% \right].
% \end{equation}

% \subsection{Reward Decomposition via Criteria-Conditioned Modeling}
\paragraph{Multi-Criteria Reward Decomposition.}
Following the intuition of attribute-conditioned modeling, we reformulate the reward function as $r_\theta(\mathcal{C}, y, \text{inst})$, where $\text{inst}$ is a criterion-specific system prompt. 
Instead of training multiple specialized models, we train a single backbone under two primary criteria:
(i) Modality-Awareness: emphasizing cross-turn acoustic coherence and prosodic naturalness. 
(ii) Colloquialness: emphasizing conversational spontaneity and the avoidance of "robotic" formalisms. This conditioned approach allows the model to share general linguistic representations while learning distinct decision boundaries for diverse evaluative dimensions, effectively replacing brittle, handcrafted rules with learnable, data-driven priors.
% \paragraph{Multi-Criteria Reward Decomposition.}
% To address the two gaps in spoken dialogue, we train the same reward model under two criteria:
% (i) a modality-aware criterion that emphasizes content adequacy, cross-turn coherence, and spoken naturalness/prosody; and
% (ii) a colloquialness criterion that emphasizes conversational spoken style.
% We implement this decomposition by conditioning the model on a criterion prompt during training and inference.
% This design allows a single backbone to share general dialogue representations while learning criterion-specific decision boundaries, and avoids brittle rule engineering for paralinguistic phenomena.
\paragraph{Loss Function.}
We optimize the model using the Bradley-Terry preference framework~\cite{19ff28b9-64f9-3656-ba40-08326a05748e}. Given a context $\mathcal{C}$ and a pair of responses $(y^+, y^-)$ where $y^+$ is preferred, the training objective is to minimize the negative log-likelihood:
\begin{equation}
    \mathcal{L}_{\text{pref}}(\theta) \! = \! -\mathbb{E}_{\mathcal{D}} \big[ \log \sigma \big( r_\theta(\mathcal{C}^+, y^+) - r_\theta(\mathcal{C}^-, y^-) \big) \big], 
\end{equation}
where $\sigma$ is the sigmoid function. This objective encourages the model to assign higher scalar rewards to responses that better satisfy the conditioned criteria within the given dialogue context.
% Since the Bradley-Terry model is inherently underdetermined—where adding a constant to all rewards does not alter preference probabilities—we incorporate a \textit{centering regularization} term $\mathcal{L}_{\text{center}}$ as proposed by \citet{eisenstein2023helping} to mitigate reward hacking and stabilize the output range:
However, strictly pairwise optimization can lead to unbounded score drift. This issue is magnified in speech reward modeling where domain shifts are prevalent. For instance, when moving from noisy YouTube audio to clean studio recordings, the model might prioritize channel characteristics as a shortcut, rewarding cleaner audio with higher absolute scores even if the dialogue quality is inferior. To address this sensitivity and stabilize the reward scale, we adopt the centering regularization term $\mathcal{L}_{center}$ from \citeposs{eisenstein2023helping}:
\begin{equation}
    \mathcal{L}_{\text{center}}(\theta) = \mathbb{E}_{\mathcal{D}} \left[ \left( r_\theta(\mathcal{C}^+, y^+) + r_\theta(\mathcal{C}^-, y^-) \right)^2 \right].
\end{equation}
The final training objective is formulated as:
\begin{equation}
    \mathcal{L}_{\text{total}}(\theta) = \mathcal{L}_{\text{pref}}(\theta) + \lambda \cdot \mathcal{L}_{\text{center}}(\theta),
\end{equation}
where $\lambda$ is the centering coefficient. 
%  This normalization ensures the rewards remain centered around zero, providing a more stable and predictable signal for subsequent reward modeling stages.
This constraint anchors the reward distribution around zero, ensuring that the model learns relative preferences within each domain rather than absolute biases based on recording conditions.

\section{Experiments}
\label{sec:exp}

% We evaluate via pairwise preference prediction on multi-turn dialogues. While operationalized as binary classification, our objective is to learn a continuous scalar reward function where score differences reflect preference probabilities (cf.\ \S\ref{sec:method}). We report pairwise accuracy on ESDR-Bench, analyzing performance consistency across different data sources (Wild/Semi-wild/Scripted) and reward criteria.

\subsection{Experiment Setup}
\label{subsec:exp_setup}

\begin{table*}[t]
\centering
\caption{\textbf{Main results on ESDR-Bench.} We report pairwise preference accuracy (\%) on the modality benchmark split into wild/semi-wild/scripted and on the colloquialness benchmark. \textbf{Modality Micro} is the weighted accuracy over all modality pairs; \textbf{Modality Macro} is the unweighted mean across the three modality subsets, serving as a stricter metric for generalization. \textbf{Overall Micro} is the weighted accuracy over all benchmark pairs, while \textbf{Overall Macro} averages modality macro and colloquialness accuracy.}
\label{tab:model-results}
\small
\begin{tabularx}{\textwidth}{l XXXXXXXX}
\toprule
\multirow{2}{*}{\textbf{Model}} & \multicolumn{3}{c}{\textbf{Modality Acc}} & \textbf{Modality} & \textbf{Modality} & \textbf{Colloq.} & \textbf{Overall} & \textbf{Overall} \\
\cmidrule(lr){2-4} \cmidrule(lr){5-6} \cmidrule(lr){7-7} \cmidrule(lr){8-9}
& Wild & Semi-wild & Scripted & \textbf{Micro} & \textbf{Macro} & \textbf{Acc} & \textbf{Micro} & \textbf{Macro} \\
\midrule
\multicolumn{9}{l}{\textit{Closed-source Judges}} \\
Gemini 2.5 Pro   & 76.12 & 65.60 & 69.77 & 72.63 & 70.50 & 98.80 & 76.42 & 84.65 \\
Gemini 2.5 Flash & 56.73 & 53.67 & 53.91 & 55.41 & 54.77 & 99.60 & 61.81 & 77.19 \\
GPT-4o Audio     & 52.61 & 49.54 & 49.26 & 51.12 & 50.47 & 98.00 & 57.91 & 74.23 \\
\midrule
\multicolumn{9}{l}{\textit{Open-source Audio LMs}} \\
Kimi-Audio        & 71.27 & 62.84 & 56.03 & 65.30 & 63.38 & 66.00 & 65.40 & 64.69 \\
Qwen 3 Omni 30B   & 64.48 & 54.59 & 48.84 & 58.18 & 55.97 & 97.20 & 63.83 & 76.59 \\
Qwen 2.5 Omni 7B  & 51.64 & 51.38 & 52.43 & 51.85 & 51.82 & 49.20 & 51.47 & 50.51 \\
Qwen 2.5 Omni 3B  & 51.88 & 44.50 & 47.15 & 49.34 & 47.84 & 52.00 & 49.73 & 49.92 \\
VITA-Audio        & 46.79 & 44.95 & 52.64 & 48.35 & 48.13 & 50.80 & 48.70 & 49.47 \\
\midrule
\multicolumn{9}{l}{\textit{Dedicated Speech Evaluators}} \\
SageLM           & 51.40 & 50.30 & 49.80 & 50.76 & 50.50 & 49.60 & 50.59 & 50.05 \\
SpeechJudge      & 54.85 & 45.70 & 57.30 & 54.44 & 52.62 & 55.20 & 54.55 & 53.91 \\
\midrule
\multicolumn{9}{l}{\textit{Cascade System}} \\
AudioReasoner+Whisper+GPT-4o & 58.62 & 47.85 & 52.79 & 55.38 & 53.09 & 75.20 & 58.25 & 64.14 \\
\midrule
\multicolumn{9}{l}{\textit{Ours}} \\
\textbf{SDiaReward 7B} & 100.00 & 92.47 & 92.27 & \textbf{96.61} & \textbf{94.91} & 97.20 & \textbf{96.70} & \textbf{96.06} \\
\textbf{SDiaReward 3B} & 99.39 & 55.38 & 82.83 & 88.62 & 79.20 & 92.00 & 89.11 & 85.60 \\
\bottomrule
\end{tabularx}
\end{table*}

% \paragraph{Datasets and Splits.}
% Spoken dialogue evaluation departs from text in two systematic ways (Figure~\ref{fig:head}). Accordingly, we construct two complementary preference subsets in SDiaReward-Dataset, each targeting a distinct gap:
% i) \textbf{Modality-gap pairs}: contrasting \emph{real conversational audio} against \emph{TTS-synthesized} speech. This subset captures paralinguistic naturalness, including prosody, emotional expressiveness, and hesitation markers. ii) \textbf{Colloquialness-gap pairs}: contrasting \emph{written-style} responses against \emph{colloquialized spoken-style} responses, targeting systematic shifts in wording and grammar.
% The colloquialness subset covers 10 domains and 25 topic scenarios. For both subsets, we split data into train/test with no overlap in dialogue episodes, preventing leakage at the multi-turn level.

\paragraph{Baselines.} We evaluate several categories of evaluators: 
1) \textbf{Zero-shot Audio Judges}, including proprietary (GPT-4o-audio~\citealp{hurst2024gpt}, Gemini 2.5~\citealp{team2023gemini}) and open-source models (Qwen-Omni~\citealp{xu2025qwen2,xu2025qwen3omnitechnicalreport}, Kimi-Audio~\citealp{ding2025kimi}, VITA-Audio~\citealp{long2025vita}).
2) \textbf{Dedicated Speech Evaluators}, including recently proposed SageLM~\cite{ge2026sagelm} and SpeechJudge~\cite{zhang2025speechjudge}.
3) \textbf{Cascade System}, implementing a pipeline of AudioReasoner~\cite{xie2025audio} + Whisper large-v3~\cite{radford2022whisper} + GPT-4o.
4) \textbf{Artifact Detection Baseline}, using Wav2Vec2-large-xlsr-deepfake~\cite{gustking2024wav2vec2} to investigate potential shortcut learning.
5) \textbf{Supervised Baselines}, specifically our SDiaReward-3B/7B, fine-tuned on \textsc{SDiaReward-Dataset} using a pairwise ranking objective via the \texttt{trl}~\cite{vonwerra2022trl} library.

% \paragraph{Baselines.} We evaluate two categories of evaluators: (1) \textbf{Zero-shot Audio Judges}, including proprietary (GPT-4o-audio~\citealp{hurst2024gpt}, Gemini 2.5~\citealp{team2023gemini}) and open-source models (Qwen-Omni~\citealp{xu2025qwen2,xu2025qwen3omnitechnicalreport}, Kimi-Audio~\citealp{ding2025kimi}, VITA-Audio~\citealp{long2025vita}). Due to the lack of existing open-source reward models specialized for multi-turn spoken dialogue, we use these generalist baselines to quantify the gap between zero-shot capabilities and targeted alignment. (2) \textbf{Supervised Baselines}, specifically our SDiaReward-3B/7B (based on Qwen2.5-Omni), fine-tuned on \textsc{SDiaReward-Dataset} using a pairwise ranking objective via the \texttt{trl}~\cite{vonwerra2022trl} library.

\paragraph{Evaluation Metrics.}
Our primary metric is \textbf{pairwise accuracy}, defined as the fraction of preference pairs whose ordering is correctly predicted by the reward scores. For a labeled preference $a \succ b$, the prediction is correct when $R_a > R_b$. To probe generalization across data regimes, we report both \textbf{Micro} and \textbf{Macro} averages. Micro accuracy aggregates over all test pairs and is dominated by larger subsets. Macro accuracy averages results over each subset, penalizing models that overfit to a single regime and providing a stricter view of generalization.

\paragraph{Implementation Details.}
Initialized from Qwen2.5-Omni, SDiaReward uses a linear head on pooled representations for scalar scoring. Audio is truncated/padded to 30s. Full hyperparameters are in Appendix~\ref{app:impl_details}.

\subsection{Main Results}
\label{sec:main_results}

Table~\ref{tab:model-results} summarizes the performance on ESDR-Bench.
\vspace{-5pt}
\paragraph{Dedicated Reward Modeling Unlocks Modality-Aware Evaluation.}
A striking observation is the struggle of general-purpose audio judges on the modality benchmark. While closed-source models like Gemini 2.5 Pro achieve saturation on colloquialness tasks ($98.80\%$), their ability to distinguish real human speech from synthesized audio is limited ($72.63\%$ Micro Acc). This suggests that zero-shot judges prioritize semantic content over acoustic naturalness. In contrast, our proposed SDiaReward-7B demonstrates substantial gains, achieving \textbf{96.61\%} Micro Accuracy on the modality benchmark. This underscores the necessity of targeted pairwise supervision for learning subtle paralinguistic preferences that general pre-training may overlook.
\vspace{-5pt}
\paragraph{The \textit{colloquialness} gap vs.\ the \textit{modality-aware} gap.}
The high performance of baseline models on the Colloquialness subset indicates that preferences for "spoken style" can often be inferred from textual/linguistic cues like grammar which are well-preserved in the semantic latent space of ALMs. However, the Modality task—requiring discrimination between two audio clips with \textit{identical} text content but differing prosody—proves much harder for baselines. SDiaReward's superior performance here confirms its ability to effectively disentangle and value acoustic nuances beyond mere semantics.

\paragraph{Performance Consistency Across Domains.}
Analysing Micro and Macro averages reveals significant differences in domain adaptability. SDiaReward-7B maintains consistent accuracy across heterogeneous splits (94.91\% Macro), mitigating the sharp divergence observed in the 3B model (88.62\% Micro vs. 79.20\% Macro). The discrepancy is most pronounced in the \textbf{Semi-wild} subset, where the 3B model's accuracy drops to $55.38\%$. This suggests that while smaller models may latch onto prominent domain features in "Wild" or "Scripted" data, the complex, "semi-scripted" nature of Semi-wild interactions requires sufficient model scale to resolve effectively.
\vspace{-2pt}
\paragraph{Human Alignment and Calibration.}
We run a blinded human study on 75 stratified pairs (Table~\ref{tab:human_eval}). Each pair is independently rated by three annotators, and we report the average agreement rate with the dataset ground-truth label. \textit{Random Sampling} shows \textbf{76.7\%} agreement, while \textit{High Confidence} \textbf{88.3\%} is higher than \textit{Low Confidence} 78.3\%, suggesting margins are indicative of human-perceived correctness. For \textit{Hard Negatives}, humans still agree with the ground truth in \textbf{93.3\%} of cases; disagreements are often from \textsc{Semi-wild (MELD)} pairs, likely related to \textbf{text--audio misalignment} and \textbf{incomplete slicing}. Overall weighted agreement is \textbf{83.5\% ($\pm$ 4.3\%)}. 
% More qualitative analysis is reported in the appendix ~\ref{sec:appendix_human}

\begin{table}[t]
\centering
\small
\caption{\label{tab:human_eval} \textbf{Human Verification Results.} We evaluate 75 stratified samples with averaged multi-annotator ratings. \textit{Human Agree.} denotes agreement with the dataset ground truth; \textit{Avg. Margin} is the model margin on each subset, and \textit{SE} reports the standard error.}
\vspace{-6pt}
\resizebox{\linewidth}{!}{
\begin{tabular}{lcccc}
\toprule
\textbf{Sample Subset} & \textbf{Count} & \textbf{Avg. Margin} & \textbf{Human Agree.} & \textbf{SE} \\
\midrule
\rowcolor{gray!10} \multicolumn{5}{l}{\textit{Validation of Model Capability}} \\
High Confidence & 20 & 1.65 & 88.3\% & $\pm$ 7.2\% \\
Low Confidence & 20 & 0.06 & 78.3\% & $\pm$ 9.2\% \\
Random Sampling & 20 & 0.77 & 76.7\% & $\pm$ 9.5\% \\
\midrule
\rowcolor{gray!10} \multicolumn{5}{l}{\textit{Error Analysis (Hard Negatives)}} \\
Model Wrong$^{\dagger}$ & 15 & -0.19 & \textbf{93.3\%} & $\pm$ 6.5\% \\
\midrule
\textbf{Overall (Weighted)} & \textbf{75} & \textbf{0.62} & \textbf{83.5\%} & \textbf{$\pm$ 4.3\%} \\
\bottomrule
\end{tabular}
}
\vspace{-4pt}
{\footnotesize $^{\dagger}$Model predicts the opposite label (mis-ranked pairs).}
\vspace{-8pt}
\end{table}
% \begin{table}[t]
% \centering
% \small
% \resizebox{\linewidth}{!}{
% \begin{tabular}{lcccc}
% \toprule
% \textbf{Sample Subset} & \textbf{Count} & \textbf{Avg. Margin} & \textbf{Human Agree.} & \textbf{Interpretation} \\
% \midrule
% \rowcolor{gray!10} \multicolumn{5}{l}{\textit{Validation of Model Capability}} \\
% High Confidence & 20 & 1.65 & \textbf{90.0\%} & Strong Alignment \\
% Low Confidence & 20 & 0.06 & 80.0\% & Calibrated Uncertainty \\
% Random Sampling & 20 & 0.77 & 75.0\% & Global Generalization \\
% \midrule
% \rowcolor{gray!10} \multicolumn{5}{l}{\textit{Validation of Data Quality (Error Analysis)}} \\
% Hard Negatives$^{\dagger}$ & 15 & -0.19 & \textbf{100.0\%} & Valid Ground Truth \\
% \bottomrule
% \end{tabular}
% }
% \caption{\label{tab:human_eval} \textbf{Human Preference Alignment.} We conducted a blind test on 75 samples. "Human Agree." denotes the agreement rate between human annotators and the dataset's ground truth labels. ($^{\dagger}$: "Hard Negatives" refers to samples where the model predicted incorrectly (negative margin). The 100\% agreement confirms that the ground truth labels were correct, and the model error stems from acoustic bias rather than label noise.)}
% \end{table}

\paragraph{Comparison with Dedicated and Cascade Evaluators.}
As shown in Table~\ref{tab:model-results}, existing dedicated speech evaluators (SageLM, SpeechJudge) hover around chance level across modality subsets, suggesting they are primarily optimized for single-turn quality rather than multi-turn conversational dynamics. Meanwhile, the Cascade system achieves strong performance on the text-driven Colloquialness task (75.20\%) but struggles significantly on Modality tasks (e.g., 47.85\% on Semi-wild). This highlights a fundamental limitation of cascade architectures: discretizing continuous audio into text inevitably loses fine-grained paralinguistic nuances and introduces cascading errors, underscoring the necessity of an end-to-end approach.

\begin{table}[t]
\centering
\small
\caption{\textbf{Modality Accuracy and Rejected Scores on OOD TTS Engines.} $S_{\text{rej}}$ denotes the average scalar reward assigned to the rejected (synthetic) audio. Higher $S_{\text{rej}}$ indicates that the synthetic speech is more challenging to distinguish from human speech.}
\label{tab:ood_results}
\resizebox{\linewidth}{!}{
\begin{tabular}{lcccccc}
\toprule
\multirow{2}{*}{\textbf{OOD Engine}} & \multicolumn{2}{c}{\textbf{Wav2Vec2-DF}} & \multicolumn{2}{c}{\textbf{SDiaReward-3B}} & \multicolumn{2}{c}{\textbf{SDiaReward-7B}} \\
\cmidrule(lr){2-3} \cmidrule(lr){4-5} \cmidrule(lr){6-7}
& Acc $\uparrow$ & $S_{\text{rej}}$ & Acc $\uparrow$ & $S_{\text{rej}}$ & Acc $\uparrow$ & $S_{\text{rej}}$ \\
\midrule
OpenAI TTS & 89.9\% & - & 93.0\% & -0.58 & \textbf{98.3\%} & -0.62 \\
CosyVoice 2 & 38.6\% & - & 93.1\% & -0.12 & \textbf{95.3\%} & -0.04 \\
FireRedTTS-2 & 94.5\% & - & 72.7\% & 0.18 & \textbf{90.9\%} & \textbf{0.29} \\
\bottomrule
\end{tabular}
}
\vspace{-8pt}
\end{table}

\paragraph{OOD Generalization and Artifact Detection.} 
To verify that SDiaReward learns true paralinguistic features rather than exploiting low-level acoustic artifacts (i.e., shortcut learning), we conduct extensive Out-of-Distribution (OOD) evaluations using three state-of-the-art TTS engines: OpenAI TTS (\texttt{gpt-4o-mini-tts})~\cite{hurst2024gpt}, CosyVoice 2~\cite{du2024cosyvoice}, and FireRedTTS-2~\cite{xie2025fireredtts}. As shown in Table~\ref{tab:ood_results}, the dedicated artifact-detection baseline Wav2Vec2-Deepfake~\cite{gustking2024wav2vec2} fails entirely against high-fidelity models like CosyVoice 2, performing below chance level (38.6\% accuracy). In contrast, SDiaReward-7B maintains robust accuracy across these unseen engines, achieving 98.3\% on OpenAI TTS, 95.3\% on CosyVoice 2, and 90.9\% on FireRedTTS-2.
Crucially, the scalar rewards assigned by SDiaReward reflect a nuanced understanding of conversational naturalness beyond binary artifact detection. FireRedTTS-2, a state-of-the-art context-aware Dialogue-TTS, receives a significantly higher average rejected score ($S_{\text{rej}} = 0.29$) compared to single-turn engines like OpenAI TTS ($-0.62$) and CosyVoice 2 ($-0.09$). This narrower margin against human speech implies that our model actively evaluates context-dependency, rightly assigning higher rewards to synthesis that exhibits superior prosodic coherence. These results empirically confirm that SDiaReward evaluates true contextual prosody rather than merely detecting low-level acoustic artifacts.

\subsection{Ablation Experiments}
\label{sec:ablation}

We conduct a comprehensive ablation study to validate our architectural choices, focusing on feature aggregation, model scaling, and loss regularization.

\vspace{-2pt}
\paragraph{Feature Aggregation and Scalability.}
We compare three pooling strategies: (1) \textit{Last Hidden State}, (2) \textit{Attention Pooling}, and (3) \textit{Mean Pooling}. As shown in Table~\ref{tab:ablation-short}, \textbf{\textit{Mean Pooling}} consistently outperforms others. We posit that while \textit{Last} pooling is sensitive to local boundary noise, \textit{Mean} pooling aggregates the holistic episode-level context, yielding a more linearly separable representation. Scaling from 3B to 7B further boosts performance, with 7B-Mean achieving state-of-the-art results.

\begin{figure*}[t]
    \centering
    \includegraphics[width=\linewidth]{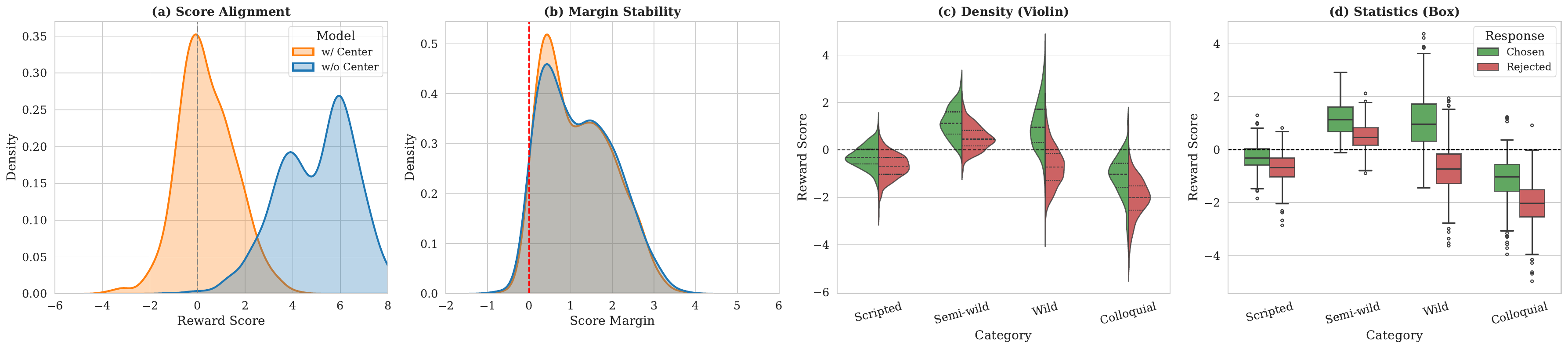}
    \vspace{-6pt}
    \caption{\textbf{Ablation Analysis on SDiaReward Model (7B).} (a) \textbf{Score Alignment}: The proposed center loss (Orange) effectively anchors the chosen reward distribution to $\mu \approx 0.32$, whereas the baseline (Blue) suffers from significant drift ($\mu > 5.0$). (b) \textbf{Margin Stability}: The discriminative margin remains robust. (c) \textbf{Density Modes}: Split violin plots visualize reward density, showing high confidence in \textit{Wild} data. (d) \textbf{Statistical Ranges}: Box plots reveal domain-dependent decision boundaries; notably, \textit{Scripted} responses receive lower absolute scores despite being correct choices.}
    \label{fig:ablation_all}
    \vspace{-8pt}
\end{figure*}

\begin{table}[t]
\centering
\small
\caption{\textbf{Ablation Study.} Performance comparison across pooling strategies and model scales. The \textit{Mean} strategy with center loss achieves the best trade-off between stability and accuracy.}
\label{tab:ablation-short}

\vspace{-6pt}
\setlength{\tabcolsep}{4pt}
\begin{tabular}{l ccc}
\toprule
\textbf{Setting} & \textbf{Modality} & \textbf{Colloq.} & \textbf{Overall} \\
\midrule
 \multicolumn{4}{l}{\textit{Pooling Strategy (3B Backbone)}} \\
Last Hidden & 63.75 & 48.80 & 61.59 \\
Attention   & 87.94 & 93.60 & 88.76 \\
\textbf{Mean} & \textbf{88.62} & \textbf{92.00} & \textbf{89.10} \\
\midrule
\multicolumn{4}{l}{\textit{Pooling Strategy (7B Backbone)}} \\
Last Hidden & 51.83 & 40.00 & 50.12 \\
Attention   & 70.60 & 55.20 & 68.37 \\
\textbf{Mean} & \textbf{96.61} & \textbf{97.20} & \textbf{96.70} \\
\midrule
\multicolumn{4}{l}{\textit{Loss Formulation (7B-Mean)}} \\
w/o Center Loss & 95.05 & 97.20 & 95.37 \\
\textbf{w/ Center Loss} & \textbf{96.61} & \textbf{97.20} & \textbf{96.70} \\
\bottomrule
\end{tabular}
\vspace{-8pt}
\end{table}

\paragraph{Impact of Center Loss Regularization.}
Standard reward modeling often suffers from unbounded score drift. Figure~\ref{fig:ablation_all}(a) illustrates this issue: the baseline model's average chosen reward drifts to $\mu \approx 5.03$. By introducing center loss, we align the global average to $\mu \approx 0.32$ (Orange curve) without compromising the discriminative margin (Fig.~\ref{fig:ablation_all}(b)). This calibration not only stabilizes training but also slightly improves accuracy ($95.37\% \to 96.70\%$) by preventing logit saturation.\looseness=-1
\vspace{-6pt}
\paragraph{Analysis of Domain-Specific Bias.}
Despite the high global accuracy, a granular analysis of score distributions reveals intrinsic domain biases. Figure~\ref{fig:ablation_all}(c) and (d) decompose the scores by data source:
\textbf{i) High Confidence in Wild Data}: The model exhibits high certainty on \textit{Wild} data, with chosen scores tightly clustered around $+0.8$ and a clear separation from rejected samples.
\textbf{ii) Adaptive Decision Boundaries}: Interestingly, for \textit{Scripted} and \textit{Colloquial} data, we observe a negative shift in the score distribution. As shown in the box plots (Figure~\ref{fig:ablation_all}(d)), the median chosen score for \textit{Scripted} data is negative ($\approx -0.24$), yet the model maintains high classification accuracy.
This phenomenon indicates that the Reward Model implicitly learns a \textit{relative} ranking function calibrated to the specific difficulty or style of each domain, rather than a globally absolute metric. While Center Loss normalizes the global mean, these local offsets suggest that future work should explore domain-invariant alignment techniques to further standardize reward scales.

\vspace{-4pt}
\section{Discussion}
\label{sec:discussion}
\vspace{-2pt}
\paragraph{The Asymmetry of Spoken Dialogue Gaps.} 
Our empirical analysis reveals a fundamental divergence where the \textit{colloquialness gap} is effectively bridged by the linguistic priors of LLMs, whereas the \textit{modality gap} remains the primary technical bottleneck. General-purpose audio models struggle to distinguish prosodic naturalness from synthesis artifacts and often perform near chance levels. \textsc{SDiaReward} resolves this by integrating modality-based supervision to ensure high-scoring responses possess both grammatical spontaneity and acoustic authenticity. This unified approach prevents the optimization pipeline from regressing into "scripted synthesis" where responses sound textually informal but prosodically rigid.

\paragraph{Reward as Relative Expressiveness.} 
\textsc{SDiaReward} goes beyond simple artifact detection to acquire a metric of \textit{relative expressiveness}. As shown in Figure~\ref{fig:ablation_all}(d), correctly ranked pairs in the \textit{Scripted} domain consistently receive lower absolute scores compared to the \textit{Wild} domain. This pattern indicates that the model implicitly calibrates to the dynamic range inherent to each domain. Such calibration is vital for reinforcement learning as it encourages the generation of emotionally rich interactive behaviors rather than spectrally clean but monotonic audio.

\paragraph{From Evaluation to Downstream Alignment.} 
While this work focuses on establishing a robust evaluation foundation, the ultimate destination of reward modeling is downstream alignment. Given the engineering complexity and computational requirements inherent in end-to-end speech generation, accurately defining, disentangling, and evaluating the modality and colloquialness gaps is a necessary prerequisite. By providing SDiaReward and ESDR-Bench, we offer a reliable "compass" for this journey. Exploring the seamless integration of our multi-criteria reward signals into downstream alignment pipelines---such as applying Direct Preference Optimization (DPO) or Group Relative Policy Optimization (GRPO) to speech-to-speech models---remains a crucial direction for future work.\looseness=-1

\vspace{-2pt}
\section{Conclusion}
\vspace{-2pt}
In this work, we take a step toward better implicitly reward modeling and evaluation for end-to-end spoken dialogue systems. We introduce \textsc{SDiaReward-Dataset}, a comprehensive pairwise preference corpus, and \textsc{ESDR-Bench} for general episode-level benchmarking. Our end-to-end reward model achieves state-of-the-art accuracy, effectively distinguishing paralinguistic naturalness and conversational spontaneity where general-purpose models fail. 
Crucially, our analysis suggests that the model learns a general measure of relative expressiveness rather than simple artifact detection. However, we also observe domain-dependent offsets in absolute reward scores. Future work should focus on deriving more general reward signals through refined data diversity and domain-invariant objectives, paving the way for stable and scalable reinforcement learning in next-generation spoken dialogue systems.\looseness=-1
% \section{Conclusion}
% \label{sec:conclusion}

% In this work, we address the core challenge of evaluating and optimizing end-to-end spoken dialogue systems by explicitly targeting the \emph{modality gap} and the \emph{colloquialness gap}. We introduce a comprehensive episode-level benchmark and a unified reward model trained on a new dataset of preference pairs. Our empirical results show that while off-the-shelf multimodal large language models struggle to distinguish human speech from synthesized audio, the proposed model achieves state-of-the-art accuracy across diverse conditions and captures both acoustic naturalness and conversational flow.
% Moreover, the score distribution analysis indicates that the model learns a robust measure of expressiveness, favoring emotionally rich interaction over monotonic speech regardless of the speech origin. By releasing the dataset and benchmark, we aim to equip the community with tools that go beyond text-based metrics. We hope this work accelerates the development of next-generation spoken dialogue systems that are not only semantically coherent, but also acoustically expressive and interactionally natural.

\section*{Limitations}
\label{sec:limitations}
While SDiaReward achieves state-of-the-art performance, our current dataset prioritizes "in-the-wild" recordings  to target the complexity of real-world acoustic environments. Future iterations could further enhance robustness by incorporating a broader spectrum of high-quality acted speech and diverse synthesis engines. Additionally, while our human verification confirms high alignment with model predictions, larger-scale studies exploring fine-grained subjective preferences remain a promising direction for future research.

\section*{Ethics and Responsible Use}
\label{sec:ethics}
This section discusses the ethical considerations, intended use, and responsible data release practices associated with SDiaReward and ESDR-Bench, with particular attention to copyright, privacy, and biometric risks in spoken dialogue research.

\paragraph{Intended Use.}
SDiaReward and ESDR-Bench are intended solely for research purposes, including the evaluation and analysis of end-to-end spoken dialogue systems and reward modeling methodologies. They are not designed for deployment in real-world decision-making systems, content moderation, surveillance, or any application involving automated judgments about individuals or groups.

\paragraph{Data Sources and Privacy.}
Our dataset is constructed from publicly available audio sources in YouTube and established research benchmarks MELD and DailyTalk. We do not redistribute raw audio recordings from third-party platforms. Our release excludes speaker-identifiable representations and persistent speaker identifiers, and provides derived research artifacts only (Appendix~\ref{app:safety}).

\paragraph{Copyright and Data Release Strategy.}
Although part of our corpus originates from publicly accessible web audio, we do not release raw audio files. To mitigate copyright risks, we release only derived artifacts such as dialogue metadata, preference annotations, benchmark splits, and evaluation scripts, strictly for non-commercial research use. Reconstructing or accessing any underlying audio content, if desired, requires users to independently obtain the data in accordance with the access conditions and platform policies of the original sources. All released resources follow the original terms and conditions of the underlying data providers.

\paragraph{Biometric and Speaker Identification Risks.}
Speech may contain biometric signals that can enable speaker identification. To reduce biometric and privacy risks, the released artifacts do not include speaker-identifiable representations and are not intended for speaker identification, biometric analysis, or any application involving individual-level profiling.

\paragraph{Risk Awareness for Downstream Optimization.}
As discussed in the Limitations section, reward models trained on heterogeneous real-world audio may exhibit sensitivity to domain-specific acoustic characteristics, which could be exploited as shortcuts during optimization. We emphasize that SDiaReward should not be treated as a substitute for human judgment and should be applied cautiously, particularly in downstream optimization settings.

\paragraph{Use of AI Assistants.}
AI assistants are used to support data preprocessing scripts and limited language refinement. All experimental design, analysis, and conclusions are determined by the authors.

% \section*{Acknowledgements}
% We sincerely thank the anonymous reviewers and meta-reviewers for their constructive comments and suggestions.

\bibliography{custom}

\appendix
% \section{Example Appendix}
% \label{sec:appendix}
% \label{app:impl_details}
\section{Training Details}
\label{app:impl_details}
We initialize SDiaReward using the Qwen2.5-Omni (3B/7B) backbone, extending it with a linear regression head atop the pooled representation of the final hidden layer to derive a scalar reward. Audio episodes are standardized to a 30-second duration via truncation or padding. The model is optimized using a Bradley-Terry pairwise loss framework, augmented with a reward centering term ($\lambda=10^{-2}$) to stabilize score distribution.Training is conducted for a single epoch using the AdamW optimizer with a peak learning rate of $2\times10^{-5}$ and a weight decay of $0.05$. We employ a cosine learning rate schedule preceded by a $0.15$ warmup phase, alongside a gradient clipping threshold of $1.0$. For computational efficiency, we leverage DeepSpeed ZeRO-2 across 4 GPUs, utilizing FlashAttention-2, \texttt{bf16} precision, and gradient checkpointing. The total batch size is configured to 32 (per-device batch size of 4 with a gradient accumulation factor of 2). Model performance is monitored every 50 steps on the validation split; the optimal checkpoint is selected based on minimal validation loss, with a rolling buffer of the 20 most recent checkpoints maintained throughout training.

\section{Reward Dataset}
\label{sec:appendix_dataset}

\subsection{Details of the Modality-aware Subset Construction}
% The SDiaReward Dataset consists of two components: Modality-gap pairs and Colloquialness-gap pairs. 

% Specifically, the Modality-gap pairs encompass three sources: Wild data, Semi-Wild data, and Scripted data, with sample sizes of 7,703, 495, and 2,658, respectively. 
\subsubsection*{Data Collection and Preprocessing}

\paragraph{Multi-source Data Acquisition Strategy}
% We adopt a hybrid data acquisition strategy that combines large-scale "in-the-wild" recordings with curated public benchmarks.
% For the large-scale "in-the-wild" data, we obtained high-quality conversation audio by selecting a group of YouTube creators who specialized in interviews and podcast production.
% We searched targeted keyword (e.g., "podcast", "interview") to identify specific content for each channel and automated the retrieval process using the \texttt{ytdlp} tool, strictly adhering to a high-fidelity protocol by selecting the best available audio streams (\texttt{bestaudio/best}) and enabling the \texttt{noclobber} parameter to ensure data integrity. This rigorous scraping pipeline yielded approximately 1,954.2 hours of raw, unconstrained audio.
% To improve generalization and mitigate overfitting, we supplemented our corpus with two authoritative public benchmarks: MELD~\cite{poria2019meld} and DailyTalk~\cite{lee2023dailytalk}.
% This combination balances the natural prosodic variability of massive unorganized audio with the structured annotations of reference datasets, creating a robust foundation for model training.
We adopt a hybrid data acquisition strategy that combines large-scale "in-the-wild" recordings with curated public benchmarks. For the large-scale "in-the-wild" data, we obtain high-quality conversation audio by selecting a group of YouTube creators who specialize in interviews and podcast production. We search targeted keyword (e.g., "podcast", "interview") to identify specific content for each channel and automate the retrieval process using the \texttt{ytdlp} tool, strictly adhering to a high-fidelity protocol by selecting the best available audio streams (\texttt{bestaudio/best}) and enabling the \texttt{noclobber} parameter to ensure data integrity. This rigorous scraping pipeline yields approximately 1,954.2 hours of raw, unconstrained audio. To improve generalization and mitigate overfitting, we supplement our corpus with two authoritative public benchmarks: MELD~\cite{poria2019meld} and DailyTalk~\cite{lee2023dailytalk}. This combination balances the natural prosodic variability of massive unorganized audio with the structured annotations of reference datasets, creating a robust foundation for model training.

\paragraph{Audio Processing Pipeline Construction}
To extract turn-level audio and its duration and text, we design a customized end-to-end processing pipeline based on the Emilia~\cite{he2024emilia} framework which handles various heterogeneous data sources. For unstructured YouTube audio, the pipeline executes a sequence of speech enhancement (\texttt{MDX23C-8KFFT-InstVoc\_HQ}\footnote{https://github.com/Anjok07/ultimatevocalremovergui}), speaker diarization (\texttt{speaker-diarization-community-1}\footnote{https://huggingface.co/pyannote/speaker-diarization-community-1}), and fine-grained VAD (\texttt{silero\_vad}~\cite{Silero_VAD}). ASR is then performed using \texttt{whisper-large-v3}~\cite{radford2022whisper}, initialized with a specific prompt to retain disfluencies (e.g., "um", "uh") and prevent filler word omission. We strictly retain only the two dominant speakers, discarding segments where secondary speakers exceed 10\% of the duration. 
For structured datasets (DailyTalk, MELD), we prioritize fidelity by bypassing VAD and ASR inference to avoid error propagation, and directly rely on the provided metadata for alignment, while applying consistent speech enhancement.
Following extraction, turn-level audio is organized into dialogue groups with granular controls: a minimum interval of 0 seconds, an overlap ratio $\ge$ 0.1, and a strict duration cap of 90 seconds. Through this rigorous pipeline, we process a total of 749.61 hours of turn-level audio from YouTube, supplemented by 21.93 hours from DailyTalk and 21.67 hours from MELD, resulting in structured JSON transcripts and segmented audio data.
% To extract turn-level audio and its duration and text, we design a customized end-to-end processing pipeline based on the Emilia~\cite{he2024emilia} framework which handles various heterogeneous data sources. For unstructured YouTube audio, the pipeline executes a sequence of speech enhancement (\texttt{MDX23C-8KFFT-InstVoc\_HQ}\footnote{https://github.com/Anjok07/ultimatevocalremovergui}), speaker diarization (\texttt{speaker-diarization-community-1}\footnote{https://huggingface.co/pyannote/speaker-diarization-community-1}), and fine-grained VAD (\texttt{silero_vad}~\cite{Silero_VAD}). ASR is then performed using \texttt{whisper-large-v3}~\cite{radford2022whisper}, initialized with a specific prompt to retain disfluencies (e.g., "um", "uh") and prevent filler word omission. We strictly retain only the two dominant speakers, discarding segments where secondary speakers exceed 10\% of the duration.
% For structured datasets (DailyTalk, MELD), we prioritize fidelity by bypassing VAD and ASR inference to avoid error propagation, and directly rely on the provided metadata for alignment, while applying consistent speech enhancement.
% Following extraction, turn-level audio is organized into dialogue groups with granular controls: a minimum interval of 0 seconds, an overlap ratio $\ge$ 0.1, and a strict duration cap of 90 seconds. Through this rigorous pipeline, we process a total of 749.61 hours of turn-level audio from YouTube, supplemented by 21.93 hours from DailyTalk and 21.67 hours from MELD, resulting in structured JSON transcripts and segmented audio data.

\paragraph{Synthetic Audio Generation and Preference Pair Organization}
% We utilized the \texttt{soulxpodcast}~\cite{SoulXPodcast} framework to generate high-quality synthetic audio for reward model training via zero-shot cloning. We designed a greedy heuristic for reference audio selection to capture rich acoustic features, prioritizing clips with a duration between 5 and 30 seconds and a word count under 60. Dialogue groups lacking viable prompts were pruned. This process yielded a cumulative synthetic corpus of 269.97 hours from YouTube, 18.97 hours from DailyTalk, and 3.58 hours from MELD. Finally, we structured the data into preference pairs: the generated synthetic audio was designated as the rejected response, while the original ground-truth audio served as the chosen response. 
We utilize the \texttt{soulxpodcast}~\cite{xie2025soulx} framework to generate high-quality synthetic audio for reward model training via zero-shot cloning. We design a greedy heuristic for reference audio selection to capture rich acoustic features, prioritizing clips with a duration between 5 and 30 seconds and a word count under 60. Dialogue groups lacking viable prompts are pruned. This process yields a cumulative synthetic corpus of 269.97 hours from YouTube, 18.97 hours from DailyTalk, and 3.58 hours from MELD. Finally, we structure the data into preference pairs: the generated synthetic audio is designated as the rejected response, while the original ground-truth audio serves as the chosen response.
% These standardized pairs were aggregated into the Hugging Face format for subsequent training stages.

\begin{table}
    \centering
    \renewcommand{\arraystretch}{0.9} 
    \scriptsize
    \setlength{\tabcolsep}{3pt} 
    \caption{Hierarchical Classification across Datasets.}
    \label{tab:taxonomy}
    \begin{tabularx}{\linewidth}{l >{\raggedright\arraybackslash}X >{\raggedright\arraybackslash}X}
        \toprule
        \textbf{Dataset} & \textbf{Primary Dimension} & \textbf{Secondary Dimension} \\
        \midrule
        \textbf{YouTube} \newline \textit{(Wild)} & 
        Anger \newline Disgust \newline Fear \newline Happiness \newline Neutral \newline Sadness \newline Surprise & 
        Cough \newline Cry \newline Filled Pauses \newline Laughter \newline Listener Feedback \newline Sigh/Breath \newline No Feature \\
        \midrule
        \textbf{MELD} \newline \textit{(Semi-wild)} & 
        Negative \newline Neutral \newline Positive & 
        Anger \newline Disgust \newline Fear \newline Joy \newline Neutral \newline Sadness \newline Surprise \\
        \midrule
        \textbf{DailyTalk} \newline \textit{(Scripted)} & 
        Commissive \newline Directive \newline Inform \newline Question \newline Unknown & 
        Anger \newline Disgust \newline Fear \newline Happiness \newline No Emotion \newline Sadness \newline Surprise \\
        \bottomrule
    \end{tabularx}
\end{table}

\begin{figure*}[t]
    \centering
    
    % --- 第一张子图 ---
    \begin{subfigure}[b]{0.23\textwidth} % 宽度设为 0.31 (约1/3)
        \centering
        \includegraphics[width=\linewidth]{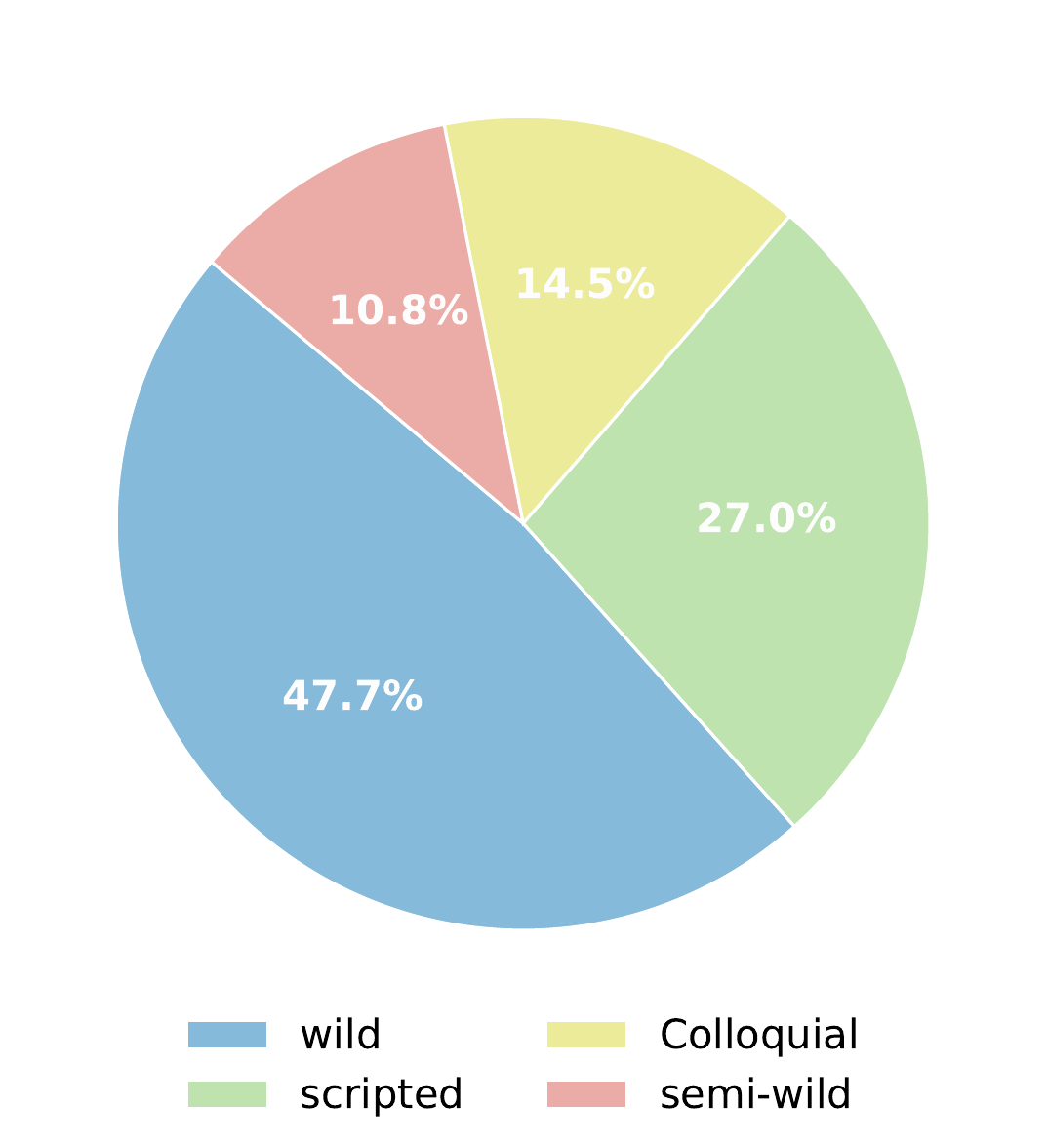}
        \caption{Category Distribution}
        \label{fig:validation_overview_1}
    \end{subfigure}
    \hfill % 【关键】在两图之间弹簧式填充空白
    % --- 第二张子图 ---
    \begin{subfigure}[b]{0.345\textwidth}
        \centering
        \includegraphics[width=\linewidth]{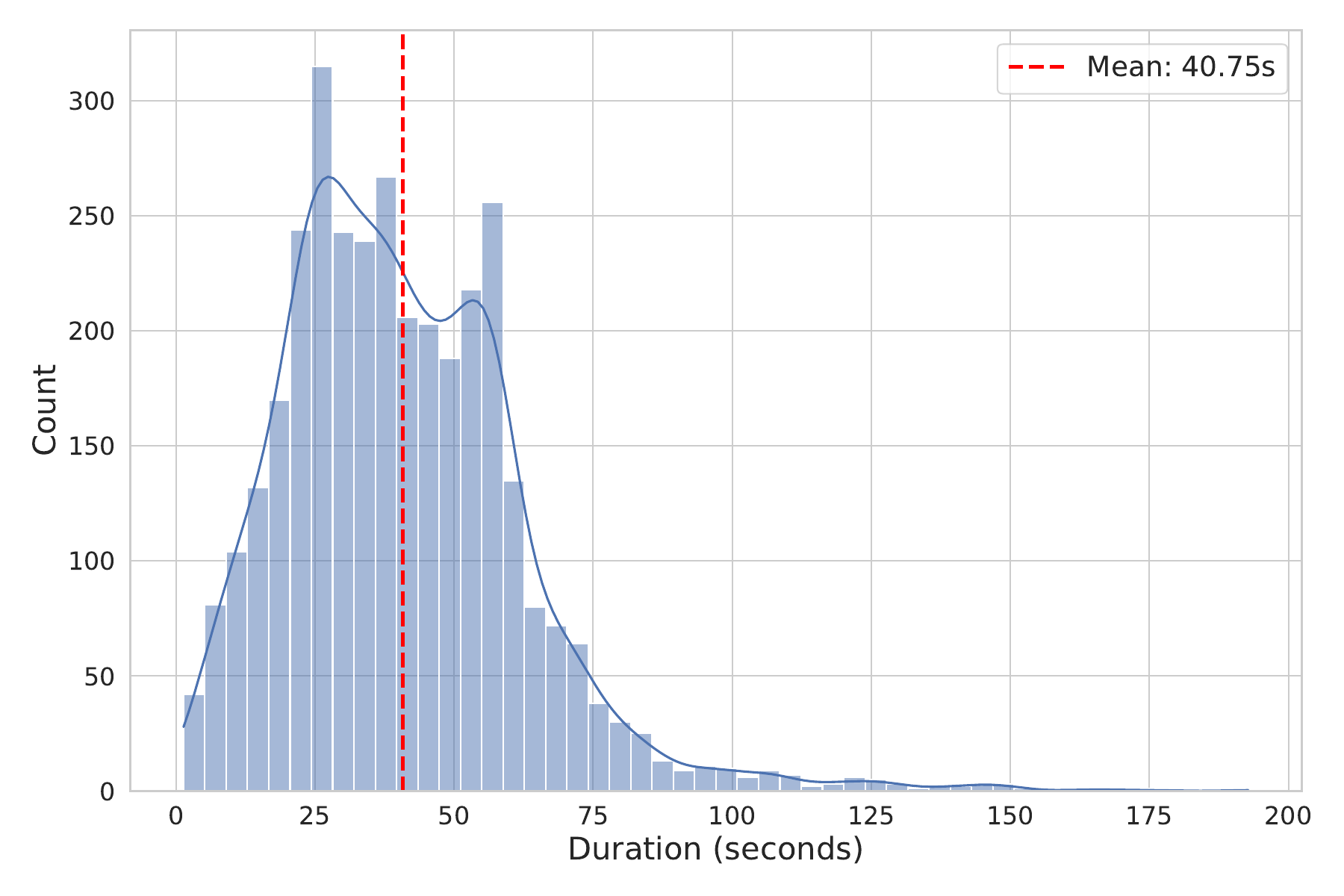}
        \caption{Duration Distribution}
        \label{fig:validation_overview_2}
    \end{subfigure}
    \hfill % 【关键】再次填充空白
    % --- 第三张子图 ---
    \begin{subfigure}[b]{0.38\textwidth}
        \centering
        \includegraphics[width=\linewidth]{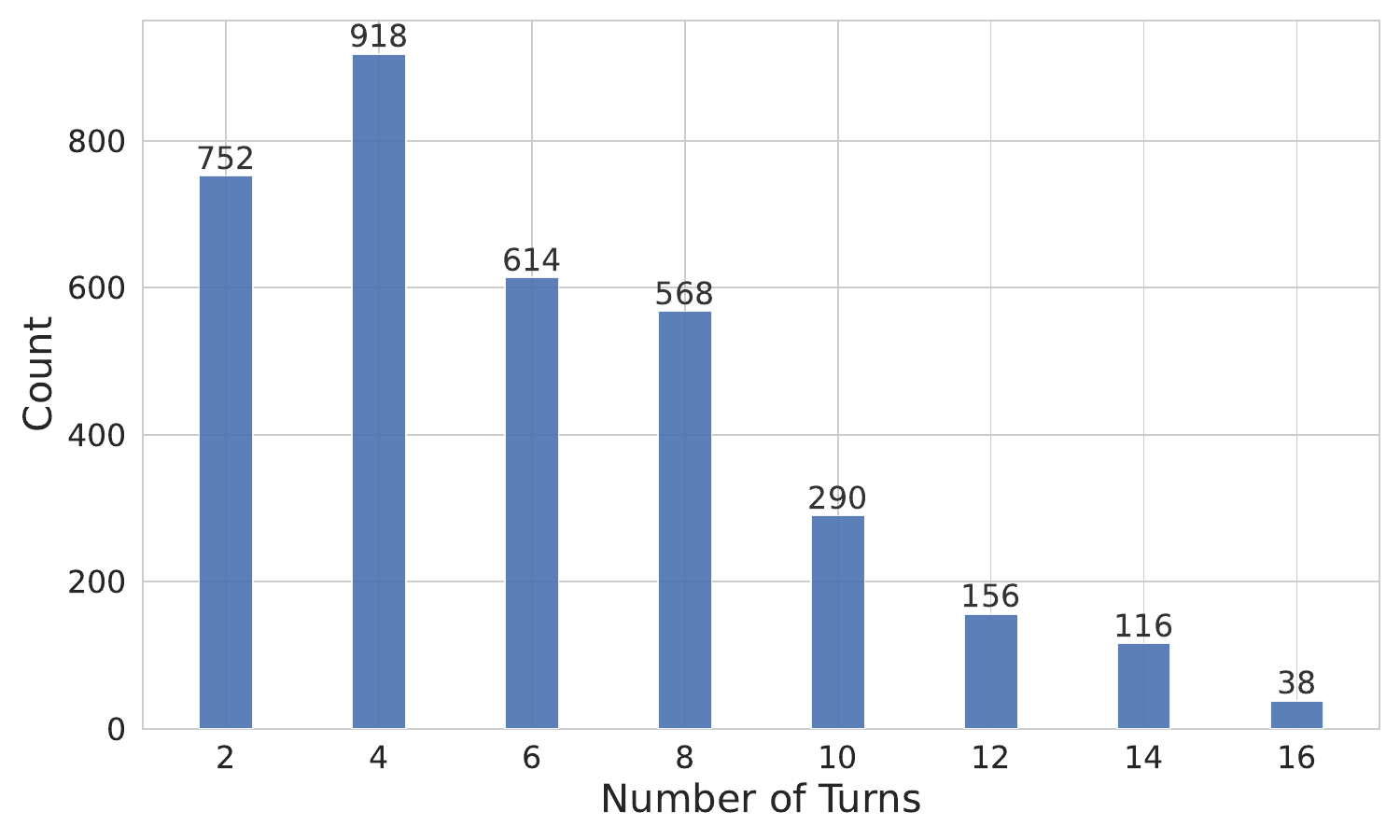}
        \caption{Turns Distribution}
        \label{fig:validation_overview_3}
    \end{subfigure}
    
    % --- 总标题 ---
    \caption{Overview of the ESDR-Bench}
    \label{fig:validation_overview}
    
\end{figure*}
\subsubsection*{Data Filtering Process and Results}
\paragraph{Phase 1: Deterministic Rule-Based Filtering}
% We applied deterministic rule-based constraints to ensure structural integrity and computational feasibility. We mandated that all dialogue groups consist of an even number of turns—guaranteeing strictly alternating user-assistant interactions—capped at a maximum of 16 turns to prevent context window overflow. Furthermore, individual turn durations were restricted to a maximum of 60 seconds. Any samples failing to meet these strict formatting or duration criteria were rigorously excised to maintain a clean and stable dataset.
We apply deterministic rule-based constraints to ensure structural integrity and computational feasibility. We mandate that all dialogue groups consist of an even number of turns—guaranteeing strictly alternating user-assistant interactions—capped at a maximum of 16 turns to prevent context window overflow. Furthermore, individual turn durations are restricted to a maximum of 60 seconds. Any samples failing to meet these strict formatting or duration criteria are rigorously excised to maintain a clean and stable dataset.

\paragraph{Phase 2: Automated Quality Assessment via LLM}
We initiate the data refinement process with an automated evaluation leveraging the multimodal capabilities of the \texttt{Gemini 2.5 Pro}~\cite{comanici2025gemini}. 
The model performs a comparative analysis between the ground-truth audio and its synthesized counterpart  within an identical multi-turn context, focusing specifically on the quality of the final turn.
The evaluation employs a 5-point scale across three dimensions: \texttt{final\_turn\_content} (semantic accuracy), \texttt{final\_turn\_naturalness\_prosody} (acoustic realism), and \texttt{dialog\_context\_coherence} (contextual logic). The model is required to output structured JSON data containing dimension-specific scores, a binary preference decision, and a concise justification ($\le$ 80 words). The specific prompt template is detailed in Figure \ref{app_fig:llm_modality_eval}.
We enforce a retention threshold based on semantic integrity; samples are preserved only if they achieve scores $\ge$ 3 in both \texttt{final\_turn\_content} and \texttt{dialog\_context\_coherence}.
% We initiate the data refinement process with an automated evaluation leveraging the multimodal capabilities of the \texttt{Gemini 2.5 Pro}~\cite{comanici2025gemini}.
% The model performs a comparative analysis between the ground-truth audio and its synthesized counterpart (designated as Version A and Version B) within an identical multi-turn context, focusing specifically on the quality of the final turn.
% The evaluation employs a 5-point scale across three dimensions: \texttt{final_turn_content} (semantic accuracy), \texttt{final_turn_naturalness_prosody} (acoustic realism), and \texttt{dialog_context_coherence} (contextual logic). The model is required to output structured JSON data containing dimension-specific scores, a binary preference decision, and a concise justification ($\le$ 80 words). The specific prompt template is detailed in Figure \ref{app_fig:data_filter}.
% We enforce a retention threshold based on semantic integrity; samples are preserved only if they achieve scores $\ge$ 3 in both \texttt{final_turn_content} and \texttt{dialog_context_coherence}.

\paragraph{Filtering Results}
% This rigorous dual-filtering mechanism effectively removed low-quality samples and disjointed contexts, thereby guaranteeing the reliability of the training corpus. The final curated dataset contains \textbf{96.48 hours} of real audio and \textbf{110.32 hours} of synthetic audio.
This rigorous dual-filtering mechanism effectively removes low-quality samples and disjointed contexts, thereby guaranteeing the reliability of the training corpus. The final curated dataset contains \textbf{96.48 hours} of real audio and \textbf{110.32 hours} of synthetic audio.

\subsubsection*{ESDR-Bench Construction}
\paragraph{Hierarchical Data Classification and Annotation}
To systematically address the heterogeneity of our data sources, we establish a hierarchical classification taxonomy tailored to the provenance of each dataset, as detailed in Table \ref{tab:taxonomy}. 
For the unstructured "Wild" data from YouTube, we employ the \texttt{Gemini 2.5 pro} to perform granular annotation, establishing emotion as the primary category and specific paralinguistic features (e.g., laughter, filled pauses) as the secondary dimension. In contrast, for the structured datasets, we prioritize fidelity to their original 
schema: MELD ("Semi-wild") is categorized primarily by sentiment followed by emotion, while DailyTalk ("Scripted") is organized by dialogue acts subdivided by emotional state.

\paragraph{Quality-Aware Sampling and Isolation}
% Based on this taxonomy, we implement a stratified sampling protocol targeting the secondary dimensions of each dataset. We applied a uniform cap to ensure balanced representation: for categories containing fewer than 50 groups, all available samples were retained; conversely, for categories exceeding this threshold, we extracted 50 instances. To guarantee a strictly independent evaluation environment, all selected validation samples were rigorously excised from the training corpus, thereby eliminating any risk of data leakage. The final resulting modality validation set comprises \textbf{14.51 hours} of real audio and \textbf{16.18 hours} of synthetic speech. 
Based on this taxonomy, we implement a stratified sampling protocol targeting the secondary dimensions of each dataset. We apply a uniform cap to ensure balanced representation: for categories containing fewer than 50 groups, all available samples are retained; conversely, for categories exceeding this threshold, we extract 50 instances. To guarantee a strictly independent evaluation environment, all selected validation samples are rigorously excised from the training corpus, thereby eliminating any risk of data leakage. The final resulting modality validation set comprises \textbf{14.51 hours} of real audio and \textbf{16.18 hours} of synthetic speech.
Figure \ref{fig:validation_overview} reports the basic statistics of the ESDR-Bench dataset.
\begin{figure*}[t]
    \centering
    
    % --- 第一张子图 ---
    \begin{subfigure}[b]{0.23\textwidth} % 宽度设为 0.31 (约1/3)
        \centering
        \includegraphics[width=\linewidth]{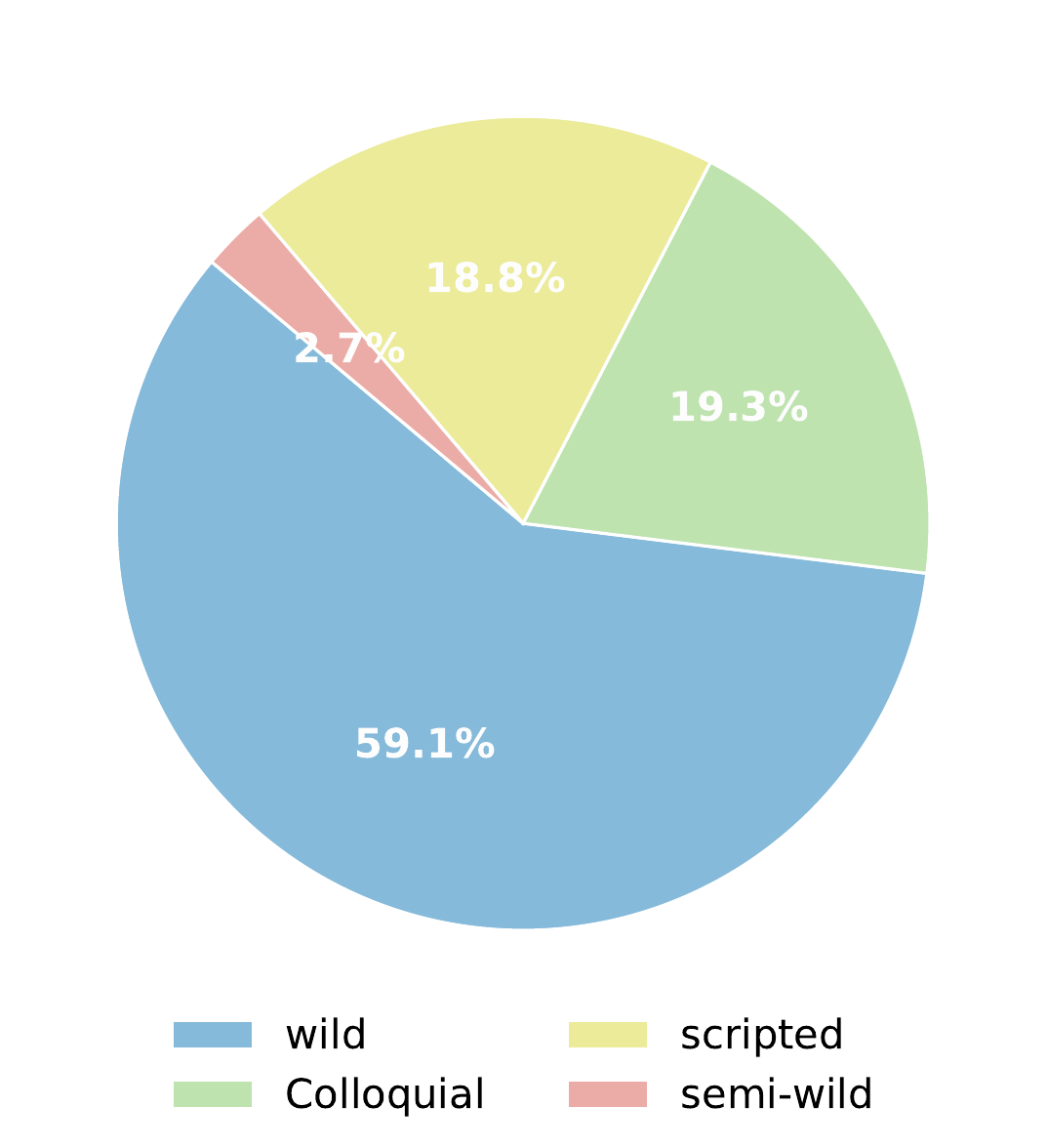}
        \caption{Category Distribution}
        \label{fig:train_overview_1}
    \end{subfigure}
    \hfill % 【关键】在两图之间弹簧式填充空白
    % --- 第二张子图 ---
    \begin{subfigure}[b]{0.345\textwidth}
        \centering
        \includegraphics[width=\linewidth]{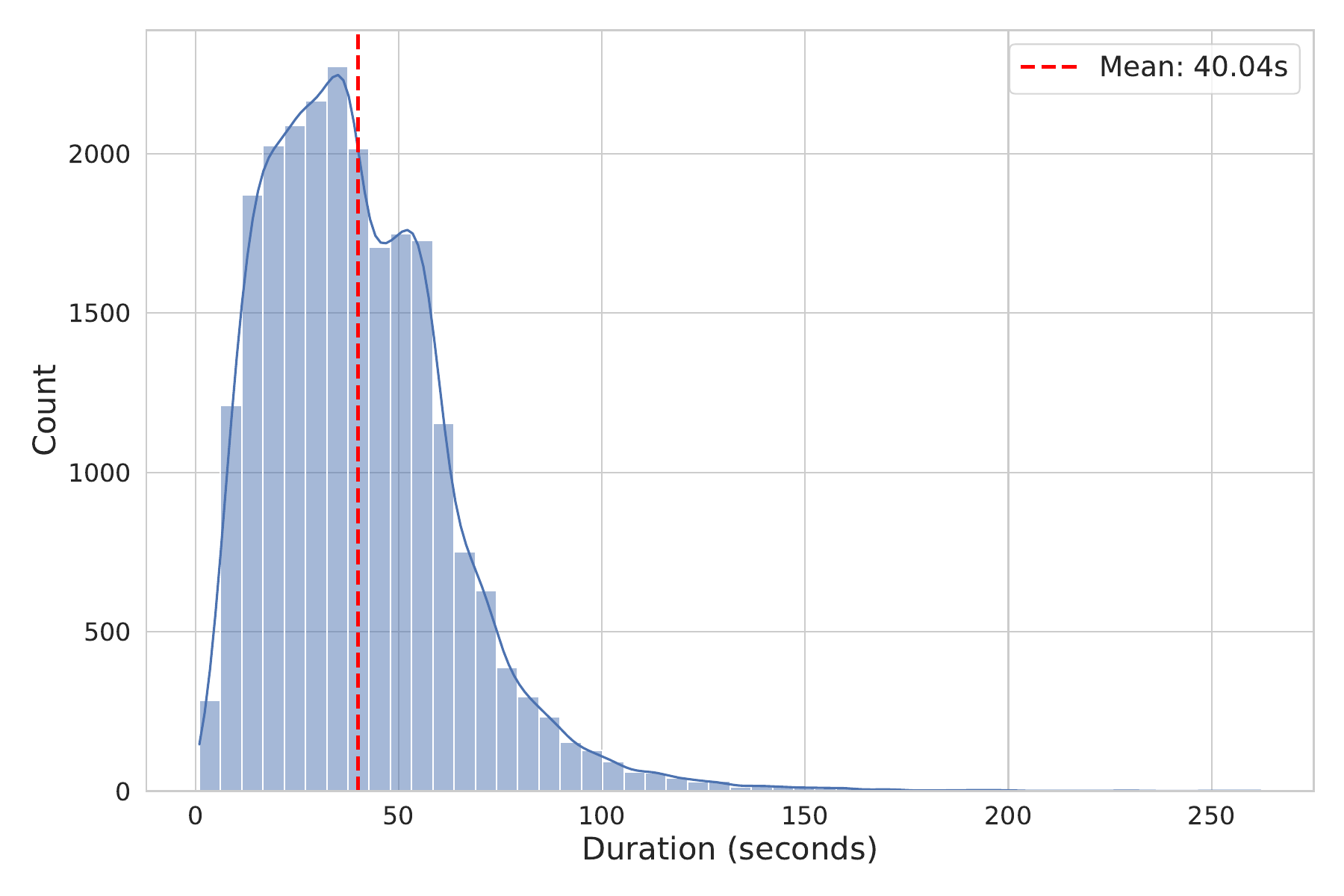}
        \caption{Duration Distribution}
        \label{fig:train_overview_2}
    \end{subfigure}
    \hfill % 【关键】再次填充空白
    % --- 第三张子图 ---
    \begin{subfigure}[b]{0.38\textwidth}
        \centering
        \includegraphics[width=\linewidth]{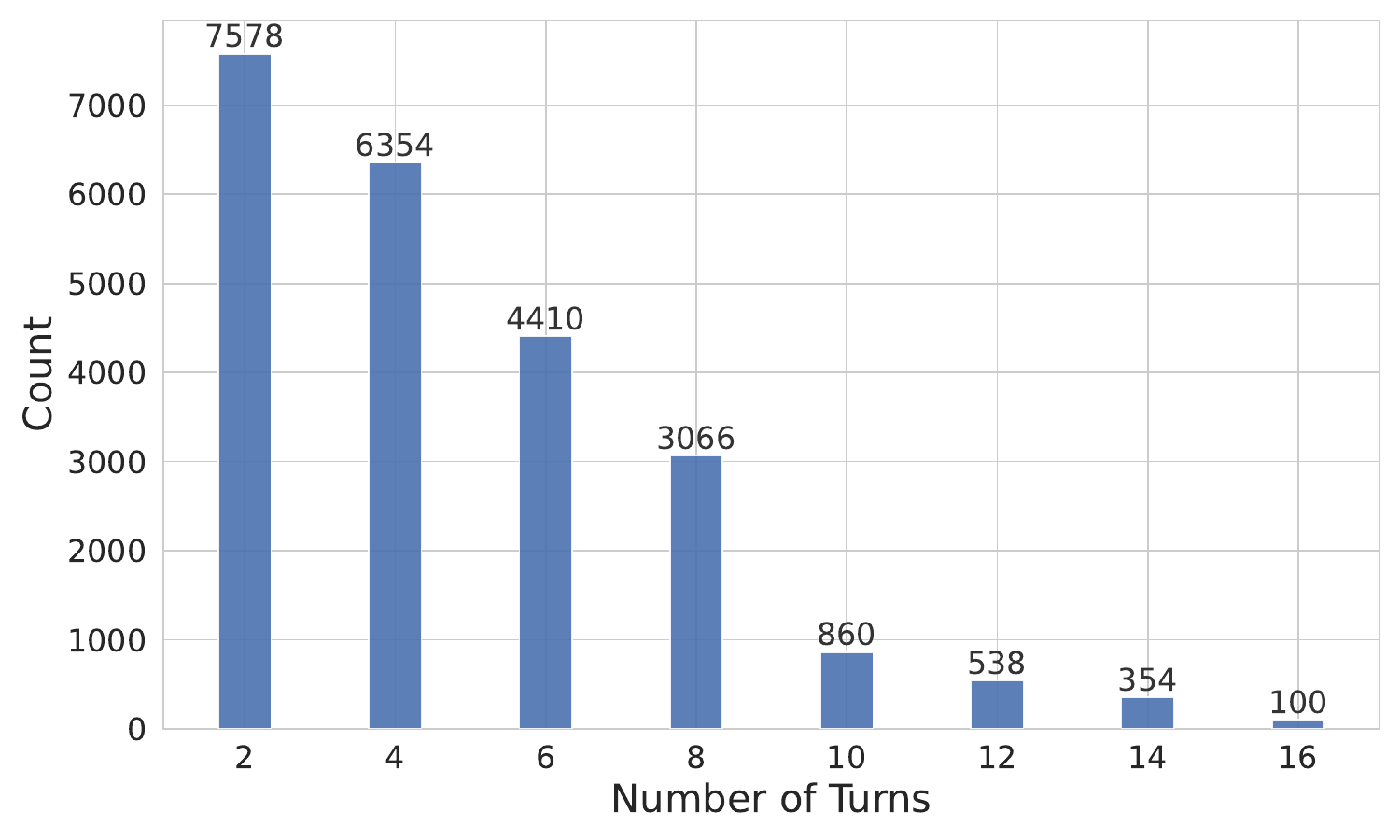}
        \caption{Turns Distribution}
        \label{fig:train_overview_3}
    \end{subfigure}
    
    % --- 总标题 ---
    \caption{Overview of the SDiaReward-Dataset}
    \label{fig:train_overview}
\end{figure*}
\subsection{Details of the Colloquialness Subset Construction}
The Colloquialness-gap pairs are constructed across ten carefully designed domains: small talk, information seeking, practical task, planning coordination, decision support, emotional support, 
\begin{figure}[htbp]
    \centering  
    \begin{subfigure}[b]{0.4\textwidth} 
        \centering
        \includegraphics[width=\linewidth]{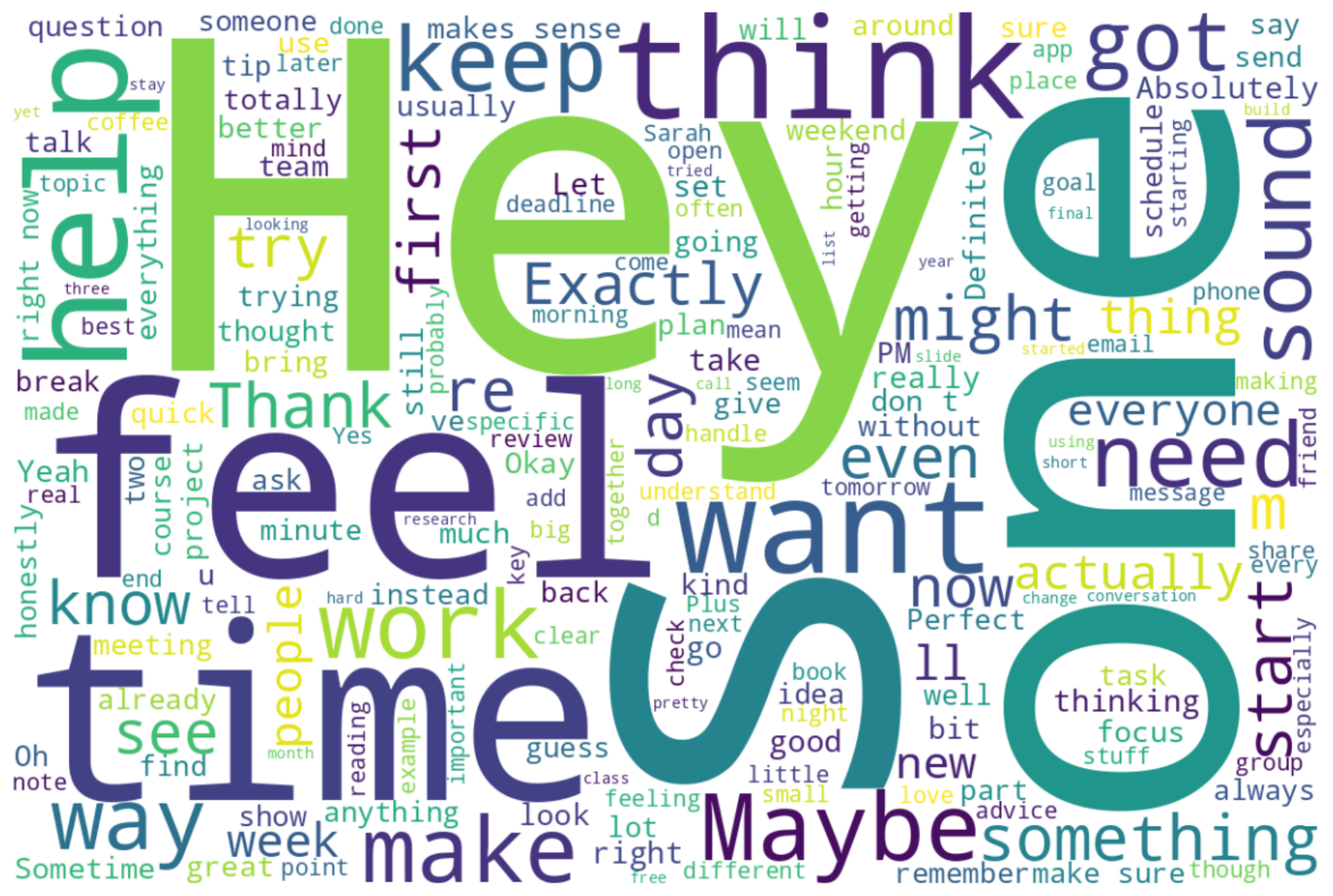} 
        \caption{Written-style Data}
        \label{fig:sub1}
    \end{subfigure}
    % \vspace{-0.1cm}
    \begin{subfigure}[b]{0.4\textwidth}
        \centering
        \includegraphics[width=\linewidth]{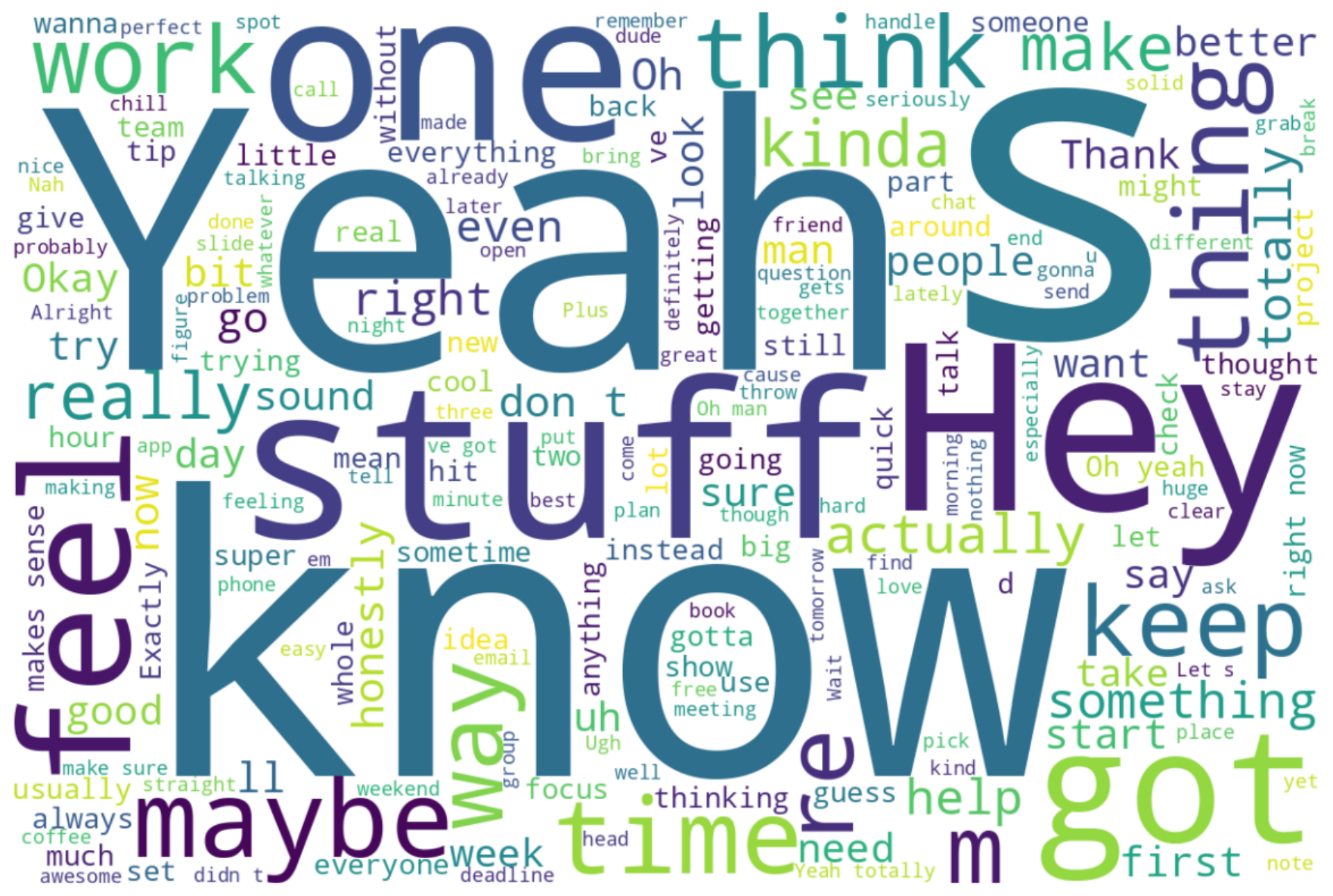}
        \caption{Spoken-style Data}
        \label{fig:sub2}
    \end{subfigure}
    \caption{Word Cloud of Colloquial Data}
    \label{fig:word_map}
\end{figure}
relationship building, social conflict, academic learning, and professional work. Within each domain, we define 25 specific topics to ensure comprehensive coverage of daily conversational scenarios. To ensure the diversity of the generated content, we utilize five models, including \texttt{Gemini 2.5 Pro}, \texttt{GPT-4.1 mini}~\cite{openai_gpt41mini_2024}, \texttt{Grok 4.1}~\cite{xai2025grok41}, \texttt{Qwen2.5-72B-Instruct}~\cite{qwen2025qwen25technicalreport} and \texttt{Qwen3-235B-Instruct}~\cite{yang2025qwen3}, to generate two written-style samples per model for each topic, resulting in a total of 2,500 samples ($10 \times 25 \times 5 \times 2$). Subsequently, these samples are rewritten into a more natural colloquial style using the aforementioned models. As illustrated in Figure~\ref{fig:word_map}, a key characteristic of the colloquial versions is the increased usage of filler words such as "yeah", "oh", and "uh". We pair the spoken-style data with the written-style data, designating the spoken version as the chosen response and the corresponding written version as the rejected response. To construct the Colloquialness subset for ESDR-Bench, we select one sample from each topic, yielding a total of 250 instances, while the remaining samples are allocated to the SDiaReward-Dataset for training purposes.

% Furthermore, SDiaReward Dataset focuses on multi-turn dialogue scenarios, with conversation lengths ranging from 4 to 16 turns; the detailed distribution is presented in Figure~\cite{?}.

% \subsection{Overview of ESDR-Bench}
% \begin{figure}[t]
%     \centering
%     \includegraphics[width=0.92\linewidth]{latex/imgs/dataset_overview.pdf}
%     \vspace{-6pt}
%     \caption{Overview of the ESDR-Bench statistics.}
%     \label{fig:bench_stat}
%     \vspace{-8pt}
% \end{figure}
% \begin{figure*}[t]
%     \centering
    
%     % --- 第一张子图 ---
%     \begin{subfigure}[b]{0.23\textwidth} % 宽度设为 0.31 (约1/3)
%         \centering
%         \includegraphics[width=\linewidth]{latex/imgs/train_label_distribution.pdf}
%         \caption{Category Distribution}
%         \label{fig:train_overview_1}
%     \end{subfigure}
%     \hfill % 【关键】在两图之间弹簧式填充空白
%     % --- 第二张子图 ---
%     \begin{subfigure}[b]{0.345\textwidth}
%         \centering
%         \includegraphics[width=\linewidth]{latex/imgs/train_duration_distribution.pdf}
%         \caption{Duration Distribution}
%         \label{fig:train_overview_2}
%     \end{subfigure}
%     \hfill % 【关键】再次填充空白
%     % --- 第三张子图 ---
%     \begin{subfigure}[b]{0.38\textwidth}
%         \centering
%         \includegraphics[width=\linewidth]{latex/imgs/train_turns_distribution.pdf}
%         \caption{Turns Distribution}
%         \label{fig:train_overview_3}
%     \end{subfigure}
    
%     % --- 总标题 ---
%     \caption{Overview of the SDiaReward-Dataset}
%     \label{fig:train_overview}
% \end{figure*}

\begin{figure*}[t]
    \centering
    \includegraphics[width=\linewidth]{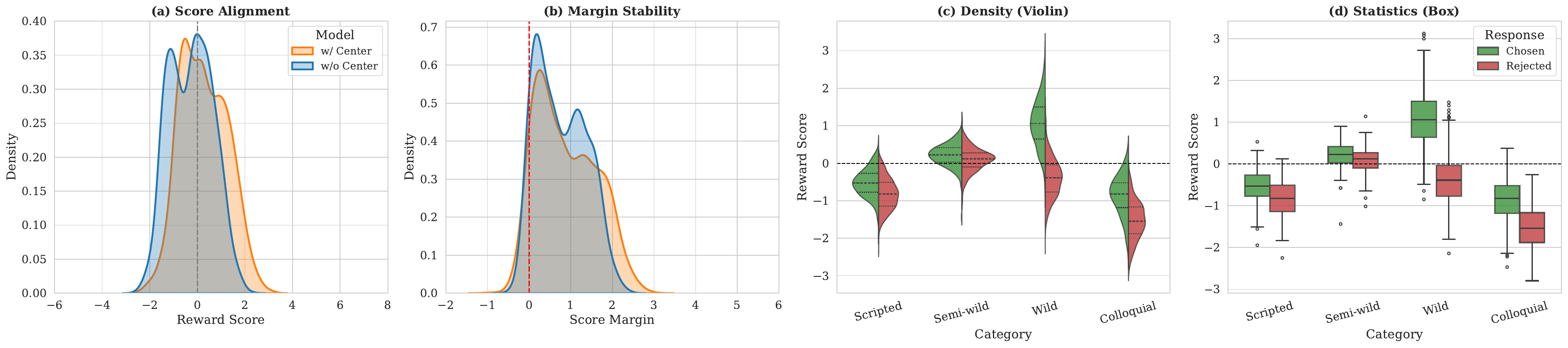}
    \vspace{-6pt}
    \caption{Ablation Analysis on SDiaReward Model (3B).}
    \label{fig:ablation_all_3B}
    \vspace{-8pt}
\end{figure*}

\subsection{Overview of the SDiaReward-Dataset}
% Having presented the data distribution of ESDR-Bench in the main text, we analyze the training subset of the SDiaReward dataset in this section. 
As shown in Figure~\ref{fig:train_overview_1}, the data is predominantly composed of "Wild" data collected from real-world scenarios (accounting for 59.1\%), while incorporating diverse categories such as "Semi-wild," "Scripted," and "Colloquial." This composition helps the model effectively narrow the modality-aware gap and the colloquialness gap. The data covers a wide distribution ranging from 2 to 16 turns, as displayed in 
Figure~\ref{fig:train_overview_2}. While dialogues with 2 to 10 turns constitute the core proportion, the substantial quantity of long-dialogue samples effectively ensures the model's ability to learn from contexts of varying lengths and grasp multi-turn interaction logic. Figure~\ref{fig:train_overview_3} indicates that the average dialogue duration is 40.04 seconds, mainly concentrated within the 20 to 80-second range. These samples possess sufficient information density and complete context, which facilitates the model in capturing richer acoustic and semantic features.

% \begin{figure}[htbp]
%     \centering
%     \includegraphics[width=0.8\linewidth]{latex/imgs/train_label_distribution.pdf}
%     \vspace{-6pt}
%     \caption{\textbf{The Category Distribution of the Training Dataset} }
%     \label{fig:train_label}
%     \vspace{-8pt}
% \end{figure}

% \begin{figure}[htbp]
%     \centering
%     \includegraphics[width=\linewidth]{latex/imgs/train_duration_distribution.pdf}
%     \vspace{-6pt}
%     \caption{\textbf{The Duration Distribution of the Training Dataset} }
%     \label{fig:train_duration}
%     \vspace{-8pt}
% \end{figure}

\subsection{Dataset Construction Prompts}
This section presents the complete prompts utilized throughout the entire construction pipeline of the SDiaReward dataset and for the evaluation of baseline models. During the dataset construction phase, Figure ~\ref{app_fig:llm_modality_eval} displays the prompt used to assess the data quality of the Modality subset to facilitate filtering and cleaning. Meanwhile, Figure ~\ref{app_fig:generate_data} and Figure ~\ref{app_fig:trans_style} present the specific instructions employed to generate the Colloquialness subset. In the model evaluation phase, we utilize the prompts shown in Figure ~\ref{app_fig:llm_modality_eval} and Figure ~\ref{app_fig:llm_colloquialness_eval}, respectively, to establish baselines for existing models regarding the Modality and Colloquialness metrics.

\subsection{Safety and Privacy Considerations in Data Release}
\label{app:safety}

Although parts of our corpus originate from publicly accessible web audio, we do not redistribute raw recordings. To reduce privacy and biometric risks, our release excludes any speaker-identifiable representations and does not provide persistent speaker-level identifiers. If transcripts are released, we remove explicit personal identifiers when detected by automatic pattern matching such as emails, phone numbers, addresses and recommend downstream users to avoid any attempt at individual-level profiling. The released artifacts are intended strictly for non-commercial research use, and derivatives of web-accessed data should not be used outside research contexts.

\begin{table*}[t]
\centering
\caption{\textbf{More results of ablation experiments}}
\label{tab:app_ablation}
\small
\begin{tabularx}{\textwidth}{l XXXXXXXX}
\toprule
\multirow{2}{*}{\textbf{Model}} & \multicolumn{3}{c}{\textbf{Modality Acc}} & \textbf{Modality} & \textbf{Modality} & \textbf{Colloq.} & \textbf{Overall} & \textbf{Overall} \\
\cmidrule(lr){2-4} \cmidrule(lr){5-6} \cmidrule(lr){7-7} \cmidrule(lr){8-9}
& Wild & Semi-wild & Scripted & \textbf{Micro} & \textbf{Macro} & \textbf{Acc} & \textbf{Micro} & \textbf{Macro} \\
% \midrule
% \textbf{Baselines} & & & & & & & & \\
% Random (20 runs) & 50.24 & 51.56 & 49.29 & 50.11 & 50.36 & 50.34 & 50.14 & 50.35 \\
\midrule
\textbf{SDiaReward 3B} & & & & & & & & \\
Last  w/o Center Loss    &66.99 	&58.06 	&39.70 	&57.25 	&54.92 	&51.20 	&56.37 	&53.06   \\
Last  w/ Center Loss     &71.60 	&55.91 	&53.00 	&63.75 	&60.17 	&48.80 	&61.59 	&54.49  \\
Atten.  w/o Center Loss    &98.91 	&55.91 	&81.33 	&87.94 	&78.72 	&93.60 	&88.76 	&86.16   \\
Atten.  w/ Center Loss    &96.72 	&77.42 	&97.64 	&94.58 	&90.59 	&65.60 	&90.38 	&78.10 \\
Mean w/o Center Loss       &100.00 	&78.49 	&80.90 	&91.26 	&86.47 	&97.60 	&92.18 	&92.03   \\
Mean w/ Center Loss       &99.39 	&55.38 	&82.83 	&88.62 	&79.20 	&92.00 	&89.11 	&85.60 
   \\

\midrule
\textbf{SDiaReward 7B} & & & & & & & & \\
Last  w/o Center Loss   &41.02 	&37.63 	&46.14 	&42.21 	&41.60 	&45.20 	&42.64 	&43.40  \\
Last  w/ Center Loss    &52.18 	&56.45 	&49.36 	&51.83 	&52.66 	&40.00 	&50.12 	&46.33  \\
Atten.  w/o Center Loss   &94.78 	&48.39 	&37.55 	&70.87 	&60.24 	&37.20 	&65.99 	&48.72    \\
Atten.  w/ Center Loss   &74.27 	&59.68 	&68.45 	&70.60 	&67.47 	&55.20 	&68.37 	&61.33  \\
Mean w/o Center Loss     &99.88 	&90.86 	&88.20 	&95.05 	&92.98 	&97.20 	&95.37 	&95.09   \\
Mean w/ Center Loss       &100.00 	&92.47 	&92.27 	&96.61 	&94.92 	&97.20 	&96.70 	&96.06 
   \\
\bottomrule
\end{tabularx}
\end{table*}
\section{Ablation Experiment}
\label{sec:appendix_ablation}
This section provides a more detailed analysis of the ablation studies. As shown in Table~\ref{tab:app_ablation}, Mean Pooling emerges as the optimal pooling strategy under both SDiaReward 3B and 7B settings, significantly outperforming the Attention Pooling and Last Hidden State strategies.

Regarding the choice of loss function, incorporating Center Loss outperforms configurations without it in the vast majority of cases. Although a slight performance decline is observed in SDiaReward 3B when combined with Mean Pooling, we ultimately adopt the scheme including Center Loss because, as illustrated in Figure ~\ref{fig:ablation_all_3B}(a), it effectively mitigates the issue of score drifting in reward modeling.

Furthermore, Figure ~\ref{fig:ablation_all_3B}(b) and Figure ~\ref{fig:ablation_all_3B}(c) reveal that SDiaReward 3B exhibits a domain-specific bias similar to that of the 7B model. This phenomenon further corroborates that the reward model implicitly learns a relative ranking function calibrated to specific domain difficulties or styles, rather than serving as a globally applicable absolute metric.

\section{Extended Discussion on Downstream Applications}
\label{sec:appendix_extended_applications}
While this work primarily focuses on the alignment and evaluation of end-to-end spoken dialogue systems, the underlying principles of our proposed \textsc{SDiaReward}---namely, capturing modality-aware paralinguistics and colloquial spontaneity---present promising avenues for various downstream applications and broader multimodal domains. Beyond unimodal speech generation, our episode-level reward framework could provide vital optimization signals for highly synchronized audio-visual generation and lip-readable speaker synthesis~\cite{yang2024synctalklip}, as well as filtering high-quality synthetic augmentations for enhancing video speech recognition~\cite{yang2024audiovsr}. Furthermore, our model's sensitivity to nuanced prosody could be extended to evaluate highly expressive vocalizations beyond standard conversational speech, such as Singing Voice Synthesis (SVS), complementing the recent curation of large-scale singing corpora~\cite{zhang2024gtsinger} and comprehensive advancements in deep-learning-based vocal synthesis~\cite{pan2025synthetic}. Finally, as foundational progress in speech recognition~\cite{EN20160220} evolves into unified, automated LLM-based generation architectures~\cite{E240342}, our framework can be integrated as an online preference judge or a dense reward signal in reinforcement learning, guiding automated agents to maintain interactional spontaneity in real-world multimodal deployments.

\begin{figure*}[t]
\centering
\begin{promptbox}
\begin{lstlisting}[
        breaklines=true,            % 自动换行，防止文字溢出
        basicstyle=\scriptsize\ttfamily, % 设置字体为打字机字体，大小为 small
        columns=fullflexible,       % 紧凑排列，避免单词间距拉大
        aboveskip=0pt,              % 去除顶部空白
        belowskip=0pt               % 去除底部空白
    ]
You are an expert in linguistics, phonetics, and conversational analysis, specializing in the intersection of multimodal prosody and spontaneous spoken language. Your mission is to evaluate two versions (A and B) of a multi-turn dialogue, focusing with surgical precision on the FINAL TURN within its context.

Evaluation Criteria
You will assess the responses based on two integrated pillars. Each pillar is composed of specific dimensions rated internally on a scale of 1 to 5.

Pillar I: Multimodal Quality & Contextual Resonance
Content Adequacy (Task Fit):

5: Perfectly captures deep intent; provides specific, forward-leaning, and pragmatically ideal responses.

3: Generally on-topic but generic; feels like a template with limited contextual depth.

1: Off-topic or ignores key constraints, appearing mismatched within the flow.

Prosodic Realism:

5: Text implies natural human rhythm and pitch; avoids a "reading aloud" feel.

3: Standard rhythm but formal/stiff; sounds like reading a script rather than talking.

1: Mechanical, flat, and devoid of emotional or rhythmic variation.

Dialogue Coherence:

5: Seamlessly integrates with the history; references or corrects details from previous turns naturally.

3: Responds to the immediate turn but lacks connection to the broader dialogue history.

1: Contextually disjointed; feels like a random insertion with no logical thread.

Pillar II: Spontaneous Colloquialness
Lexical Informality:

5: Uses authentic everyday language and idioms; completely avoids "corporate" or formal tones.

1: Highly formal, filled with jargon or "written-only" vocabulary.

Syntactic Fragmentation:

5: Features natural pauses, short clauses, and self-repairs typical of "thinking while speaking."

1: Perfect, complex textbook grammar that feels unnatural for spontaneous speech.

Spoken Markers & Disfluencies:

5: Naturally embeds fillers (um, uh, well) and particles that enhance realism.

1: Sterilized of all markers; sounds like a perfectly edited text file.

Pragmatic Naturalness:

5: High interpersonal warmth; displays empathy, humor, or hesitation as a human would.

1: Operates as a cold information retrieval machine without social nuance.

Textual Prosody:

5: Masterful use of punctuation and particles (e.g., "right?", "...", "!") to convey mood and stress.

1: Rigid or absent punctuation; the text feels lifeless and cold.

Strict Output Requirements
Your decision must prioritize the version that sounds most like a living person in a real-world setting.

Output format MUST be exactly two blocks:

<think> the reasoning progress here</think>

<answer> A or B </answer>
\end{lstlisting}
\end{promptbox}
\vspace{-2pt}
\caption{Prompt for Data Filtering.}
\label{app_fig:data_filter}
\end{figure*}

\begin{figure*}[t]
\centering
\begin{promptbox}
\begin{lstlisting}[
        breaklines=true,            % 自动换行，防止文字溢出
        basicstyle=\scriptsize\ttfamily, % 设置字体为打字机字体，大小为 small
        columns=fullflexible,       % 紧凑排列，避免单词间距拉大
        aboveskip=0pt,              % 去除顶部空白
        belowskip=0pt               % 去除底部空白
    ]
You are a data generator for an English dialogue dataset.

Your job:
- Create natural dialogues between "user" and "assistant".
- The conversation must contain between 4 and 16 rounds.
- Each round consists of exactly one "user" message followed by one "assistant" message.
- All content must be in ENGLISH.
- The dialogue must be realistic and coherent.
- Do NOT add any explanations, comments or code fences.
- Output MUST be a JSON array only.
- Do NOT use any emojis or emoticons.
- Use rich and varied sentence structures and vocabulary; avoid repetitive wording and patterns.
\end{lstlisting}
\end{promptbox}
\begin{user_promptbox}
\begin{lstlisting}[
        breaklines=true,            % 自动换行，防止文字溢出
        basicstyle=\scriptsize\ttfamily, % 设置字体为打字机字体，大小为 small
        columns=fullflexible,       % 紧凑排列，避免单词间距拉大
        aboveskip=0pt,              % 去除顶部空白
        belowskip=0pt               % 去除底部空白
    ]
You are generating variation {sample_idx + 1} of {total_samples} for the topic: "{topic}".

Requirements:
- Style: {style_desc}
- This variation MUST be clearly different from other possible dialogues on the same topic:
  - Use a different concrete situation (time, place, details).
  - Use different wording, phrases, and conversation structure.
  - Change the characters' personalities or attitudes if reasonable.
- The conversation must contain between 4 and 16 rounds.
- Each round consists of exactly one "user" message followed by one "assistant" message.
- Roles are only "user" and "assistant".
- The first message MUST be from the "user".
- The last message MUST be from the "assistant".
- Roles must strictly alternate between "user" and "assistant".
- Output format:
  A JSON array, e.g.
  [
    {{"role": "user", "content": "..." }},
    {{"role": "assistant", "content": "..." }}
  ]

Important:
- Do NOT wrap the JSON in any markdown code block.
- Do NOT include any text before or after the JSON.
- Output ONLY the JSON array.
\end{lstlisting}
\end{user_promptbox}
\vspace{-2pt}
\caption{Prompt for Written-style Data.}
\label{app_fig:generate_data}
\end{figure*}

\begin{figure*}[t]
\centering
\begin{promptbox}
\begin{lstlisting}[
        breaklines=true,            % 自动换行，防止文字溢出
        basicstyle=\scriptsize\ttfamily, % 设置字体为打字机字体，大小为 small
        columns=fullflexible,       % 紧凑排列，避免单词间距拉大
        aboveskip=0pt,              % 去除顶部空白
        belowskip=0pt               % 去除底部空白
    ]
You are a dialogue style converter.
Goal:
Convert "written" dialogues into more casual, highly colloquial "spoken" English while keeping their original meaning and information intact.
Input:
* A JSON object with fields:

  * "domain" (string)
  * "topic" (string)
  * "style" (string, currently "written")
  * "sample_id" (number)
  * "dialogue": a list of turns; each turn has:
    * "role": "user" or "assistant"
    * "content": string
Output:
* A JSON object with EXACTLY the same schema and fields:

  * Keep "domain", "topic", and "sample_id" unchanged.
  * Set "style" to "spoken".
  * "dialogue" must keep:

    * the same number of turns
    * the same order of turns
    * the same "role" values
      but rewrite each "content" into highly colloquial, natural spoken English.
Requirements:
* Preserve the original meaning and information of every utterance.
* Make the style very conversational, reflecting spontaneous spoken language:
  * Use informal, everyday words instead of formal or complex written expressions.
  * Employ short sentences, fragments, and incomplete thoughts that are typical of natural speech.
  * Incorporate fillers like "um", "uh", "you know", "I mean", and "like" naturally and sparsely.
  * Add short interjections like "yeah", "oh", "wow", and "hey" to make the conversation flow more naturally.
  * Include hesitation markers, repetition, self-repairs, and backchannels to make the dialogue feel more real, as if it's happening in real-time.
  * Use punctuation that reflects the natural rhythm of speech, such as commas, ellipses, or exclamation marks to represent pauses, emphasis, or excitement.
  * Make the dialogue feel interpersonal, casual, and contextually appropriate, just like an interaction between two people.
* Do NOT use emojis or any kind of internet slang.
* Do NOT change the language (keep it in English).
* Do NOT add or remove dialogue turns.
* Do NOT swap the roles of "user" and "assistant".
* Do NOT add explanations, comments, or extra keys.
Output format:
* Return ONE single JSON object only.
* NO markdown, NO code fences, NO extra commentary.
\end{lstlisting}
\end{promptbox}
\vspace{-2pt}
\caption{Prompt for Spoken-style Data.}
\label{app_fig:trans_style}
\end{figure*}

\begin{figure*}[t]
\centering
\begin{promptbox}
\begin{lstlisting}[
        breaklines=true,            % 自动换行，防止文字溢出
        basicstyle=\scriptsize\ttfamily, % 设置字体为打字机字体，大小为 small
        columns=fullflexible,       % 紧凑排列，避免单词间距拉大
        aboveskip=0pt,              % 去除顶部空白
        belowskip=0pt               % 去除底部空白
    ]
You are an expert in linguistics, phonetics, and conversational analysis.

You will listen to two versions of the same multi-turn dialogue, each containing several context turns and a final turn.
We are especially interested in the quality of the FINAL TURN of each version, evaluated in the context of the whole dialogue.

Your task is: after fully understanding the preceding context, evaluate the FINAL TURN of version A and version B from the following three dimensions. For each dimension, you must assign an integer score from 1 to 5 for both A and B, and then give an overall preference between A and B.

[Dimension 1: Final-turn content quality and task adequacy (final_turn_content)]
Focus: whether the final turn, given the previous dialogue context, is appropriate, relevant, and useful.

- 5: Excellent. Accurately understands the prior context and user intent, directly addresses the core issue, provides specific and helpful information, and pragmatically fits the situation, naturally advancing or closing the dialogue.
- 4: Good. Mostly understands the context, stays on-topic, gives reasonably helpful content, and is pragmatically appropriate.
- 3: Fair. Not clearly off-topic, but content is generic, shallow, or template-like, with limited usefulness.
- 2: Poor. Partially off-topic or ignores key context; content is repetitive, empty, or pragmatically somewhat inappropriate.
- 1: Very poor. Severely mismatched with the context, possibly contradictory or strongly inappropriate in pragmatics.

[Dimension 2: Final-turn spoken naturalness and prosodic realism (final_turn_naturalness_prosody)]
Focus: how the final turn is spoken - does it sound like a human naturally speaking in this situation?

- 5: Very high naturalness. Almost indistinguishable from real human speech; natural pitch and rhythm, appropriate pauses and emotions, no obvious synthetic or "reading aloud" impression.
- 4: High naturalness. Generally natural, with some prosodic variation and reasonable pauses; synthetic or "TTS-like" artifacts are minor.
- 3: Medium naturalness. Some spoken flavor and prosody, but elatively regular or "read-aloud" in style; may sound like careful reading rather than casual conversation.
- 2: Low naturalness. Clear TTS/robotic or "reading" impression; flat prosody, stiff pauses, weak emotion.
- 1: Very low naturalness. Strongly mechanical or synthetic, almost no natural prosody or emotion.

[Dimension 3: Dialogue-level coherence and context utilization (dialog_context_coherence)]
Focus: how well the final turn makes use of the prior dialogue context and fits into the multi-turn interaction.

- 5: Excellent. Clearly grounded in specific details from the previous turns (e.g., referencing, following up, correcting, or responding to emotions); the dialogue progression feels natural and coherent.
- 4: Good. Clearly responds to the right speaker and topic; thematically aligned and reasonably coherent with the context.
- 3: Fair. Topic is not completely off, but the connection to context is weak; feels more like single-turn QA than true dialogue.
- 2: Poor. Weak contextual linkage; transitions feel abrupt or template-like; limited use of prior context.
- 1: Very poor. Almost ignores the previous dialogue; off-topic or contradictory; feels like a random line inserted into the dialogue.

You MUST use only the integers 1, 2, 3, 4, or 5 for all scores.

Please output your result in the following JSON format (and DO NOT include any extra text outside the JSON):

{
  "scores": {
    "final_turn_content": { "A": <1-5>, "B": <1-5> },
    "final_turn_naturalness_prosody": { "A": <1-5>, "B": <1-5> },
    "dialog_context_coherence": { "A": <1-5>, "B": <1-5> }
  },
  "overall_preference": "A" or "B",
  "reason": "Use 2-4 sentences to briefly explain which version you prefer and why, in no more than 80 words."
}
\end{lstlisting}
\end{promptbox}
\vspace{-2pt}
\caption{Prompt for Modality Evaluation.}
\label{app_fig:llm_modality_eval}
\end{figure*}

\begin{figure*}[t]
\centering
\begin{promptbox}
\begin{lstlisting}[
        breaklines=true,            % 自动换行，防止文字溢出
        basicstyle=\scriptsize\ttfamily, % 设置字体为打字机字体，大小为 small
        columns=fullflexible,       % 紧凑排列，避免单词间距拉大
        aboveskip=0pt,              % 去除顶部空白
        belowskip=0pt               % 去除底部空白
    ]
You are an expert in linguistics, phonetics, and conversational analysis.

You will evaluate two versions of the same multi-turn dialogue. Each version has several context turns and a final turn. Your task is to evaluate both versions based on their **colloquialness**.

Colloquialness refers to the degree to which an utterance resembles spontaneous spoken language rather than formal written or scripted text. It is a multidimensional property characterized by the following criteria:

1. **Lexical informality** - Use of everyday, casual words instead of formal written expressions.
2. **Syntactic simplicity & fragmentation** - Shorter clauses, incomplete sentences, self-repairs, and other structures typical of spoken production.
3. **Presence of disfluencies and discourse markers** - Fillers ("um", "uh", "em"), hesitation markers, repetition, backchannels, and interactional cues.
4. **Pragmatic naturalness** - Contextually appropriate, human-like, interpersonal, and interaction-driven expressions.
5. **Textually implied prosody** - Usage of punctuation, particles, and exclamations that reflect spoken rhythm, emphasis, and emotional nuance.

**Your task is to evaluate both versions of the dialogue according to the above criteria**. For each dimension, you must assign an integer score from 1 to 5 for both versions (A and B) and provide an overall preference.

### Scoring:
- **5**: Very high colloquialness. The dialogue sounds highly natural, as if spontaneously spoken, with all the features of spoken language.
- **4**: High colloquialness. The dialogue is mostly casual and spontaneous, with some elements of spoken language, though minor parts may still resemble written or edited speech.
- **3**: Medium colloquialness. The dialogue includes some colloquial features, but there are noticeable elements that sound more scripted or formal.
- **2**: Low colloquialness. The dialogue is mostly formal, structured, or edited, with few to no features of natural speech.
- **1**: Very low colloquialness. The dialogue is highly formal, written, or scripted, lacking any signs of spontaneous, natural speech.

### Evaluation:
For each version, you will give a score for each of the following dimensions:
- **Lexical informality**
- **Syntactic simplicity & fragmentation**
- **Presence of disfluencies and discourse markers**
- **Pragmatic naturalness**
- **Textually implied prosody**

After assigning scores for all the dimensions, you must also provide an overall preference between A and B.

Please output your result in the following JSON format (and DO NOT include any extra text outside the JSON):

{
  "scores": {
    "lexical_informality": { "A": <1-5>, "B": <1-5> },
    "syntactic_simplicity": { "A": <1-5>, "B": <1-5> },
    "disfluencies": { "A": <1-5>, "B": <1-5> },
    "pragmatic_naturalness": { "A": <1-5>, "B": <1-5> },
    "prosody": { "A": <1-5>, "B": <1-5> }
  },
  "overall_preference": "A" or "B",
  "reason": "Use 2-4 sentences to briefly explain which version you prefer and why, in no more than 80 words."
}
\end{lstlisting}
\end{promptbox}
\vspace{-2pt}
\caption{Prompt for Colloquialness Evaluation.}
\label{app_fig:llm_colloquialness_eval}
\end{figure*}

\end{document}